\documentclass[english,twocolumn,prl,preprintnumbers,amsmath,amssymb,superscriptaddress]{revtex4}
\usepackage{verbatim}
\usepackage{graphicx}
\usepackage{amssymb}
\usepackage{babel}
\usepackage{float}
\usepackage{titlesec}
\makeatother

\newcommand{\RN}[1]{%
	\textup{\uppercase\expandafter{\romannumeral#1}}
}
\begin{document}
	
	\preprint{XXX}

	\title{Giant Tunable Mechanical Nonlinearity in Graphene-Silicon Nitride Hybrid Resonator}
	
	\author{Rajan Singh}
	\affiliation{Department of Physics, Indian Institute of Technology - Kanpur, UP-208016, India}
	\author{Arnab Sarkar}
	\affiliation{Department of Physics, Indian Institute of Technology - Kanpur, UP-208016, India}
	\author{Chitres Guria}
	\affiliation{Department of Physics, Indian Institute of Technology - Kanpur, UP-208016, India}
	
	\author{Ryan J.T. Nicholl}
	\affiliation{Department of Physics and Astronomy, Vanderbilt University, Nashville, Tennessee 37235, USA}
	\author{Sagar Chakraborty}
	\affiliation{Department of Physics, Indian Institute of Technology - Kanpur, UP-208016, India}

	\author{Kirill I. Bolotin}
	\affiliation{Department of Physics, Freie Universitat Berlin, Arnimallee 14, Berlin 14195, Germany}
	
	\author{Saikat Ghosh}
	\email{gsaikat@iitk.ac.in}
	\affiliation{Department of Physics, Indian Institute of Technology - Kanpur, UP-208016, India}



	\begin{abstract}
	
	\textbf{High quality factor mechanical resonators have shown great promise in developing classical or quantum technologies. Simultaneously, progress has been made in developing controlled mechanical nonlinearity. Here we combine these two directions of progress in a single platform consisting of coupled Silicon Nitride (SiNx) and graphene mechanical resonators. We show that nonlinear response can be induced on a large area SiNx resonator mode and can be efficiently controlled by coupling it to a gate-tunable, freely suspended graphene mode. The induced nonlinear response of the hybrid modes, as measured on the SiNx resonator surface is giant, with one of the highest measured Duffing constants. We observe a novel phononic frequency comb which we use as an alternate validation of the measured values, along with numerical simulations which are in overall agreement with measurements.}

\end{abstract}

\maketitle

\section*{Introduction}
For more than a century, mechanical resonators~\cite{Braginsky92}  have played a central role in measuring forces~\cite{Cavendish-1798,Abbott-16,Weber-16} and testing fundamental physical principles~\cite{Clerk10,Wollman-15,Ockeloen-Korppi-18,Marinkovic-18}. With the advent of micro and nanoscale mechanical resonators, and in particular, after experimental observation of their quantum mechanical behavior~\cite{OConnell-10,Taufel-11,Schliesser-11,Aspelmeyer-14}, there has been a renewed interest in usage of such resonator modes in classical~\cite{Mahboob-11} and quantum technologies~\cite{Rips-16,Mika-07}. Significant progress has been made in two broad directions over the last decade. On one hand, there has been progress in developing high quality factor ($Q$) mechanical resonators modes at high resonant frequency ($f_0$) towards quantum devices at room temperature~~\cite {kippenberg-07, chakram-14}. Silicon Nitride (SiNx) has emerged as a dominant material of choice~\cite{Fink-16} for such resonators, demonstrating mechanical $Q$'s in excess $10^7$ at MHz frequencies~\cite{Reinhardt-16,Norte-16,Patil-15}. On the other hand, progress has been made in developing tunable mechanical nonlinearity, towards conditional phase shifts of mechanical modes~\cite{Mahboob-11}. Freely suspended graphene resonator, with low mass and high Young's modulus~\cite{Lee-08,Eichler-11,Davidovikj-17,Storch-18} resulting in exceptional nonlinear elastic properties~\cite{Eichler-11,Davidovikj-17} along with gate tunable resonant frequency, emerged as an efficient choice;  mixing~\cite{Karabalin-09,Mahboob-12,Cao-14,Mahboob-16,Ganesan-17,Seitner-17,Ganesan-18,Czaplewski-18} and side-band cooling of its strongly coupled modes have been observed~\cite{Alba-16,Mathew-16}. A platform that can integrate these two directions of progress, combining high quality factors of SiNx resonators with gate tunable response of graphene resonators can be a logical next step~\cite{Fink-16}.

Here we explore a hybrid platform consisting of a large area SiNx resonator coupled to an atomically thin, freely suspended graphene that is deposited on holes etched on the SiNx. When a mechanical mode of the graphene resonator is electrostatically tuned into resonance with a SiNx mechanical mode, we observe the resulting hybrid modes develop giant nonlinear response to an external driving force, as measured on the surface of the SiNx resonator (fig.~1a). To validate the measurements, we develop an alternate, novel methodology to characterize third order (Duffing) nonlinear response and damping coefficients of these hybrid modes. By parametrically driving the coupled modes, we observe generation of novel frequency comb and use the measured amplitudes of the generated comb lines to estimate the nonlinearities and validate the results. Our measurements are in agreement with numerical simulations of a model of coupled linear and nonlinear oscillators. The model suggests that induced nonlinearity of hybrid modes, as measured on SiNx, is due to back-action force of the nonlinear graphene resonator and simple scaling estimates are in agreement with the measured giant values. These result, verified with two separate measurements, thereby combine two directions of development of electromechanical resonators in a single hybrid device -- a gate tunable graphene resonator inducing giant nonlinearity in a high-$Q$ mechanical mode of a large area SiNx resonator (fig. 1a).\\



\begin{figure*}\includegraphics[scale=0.6]{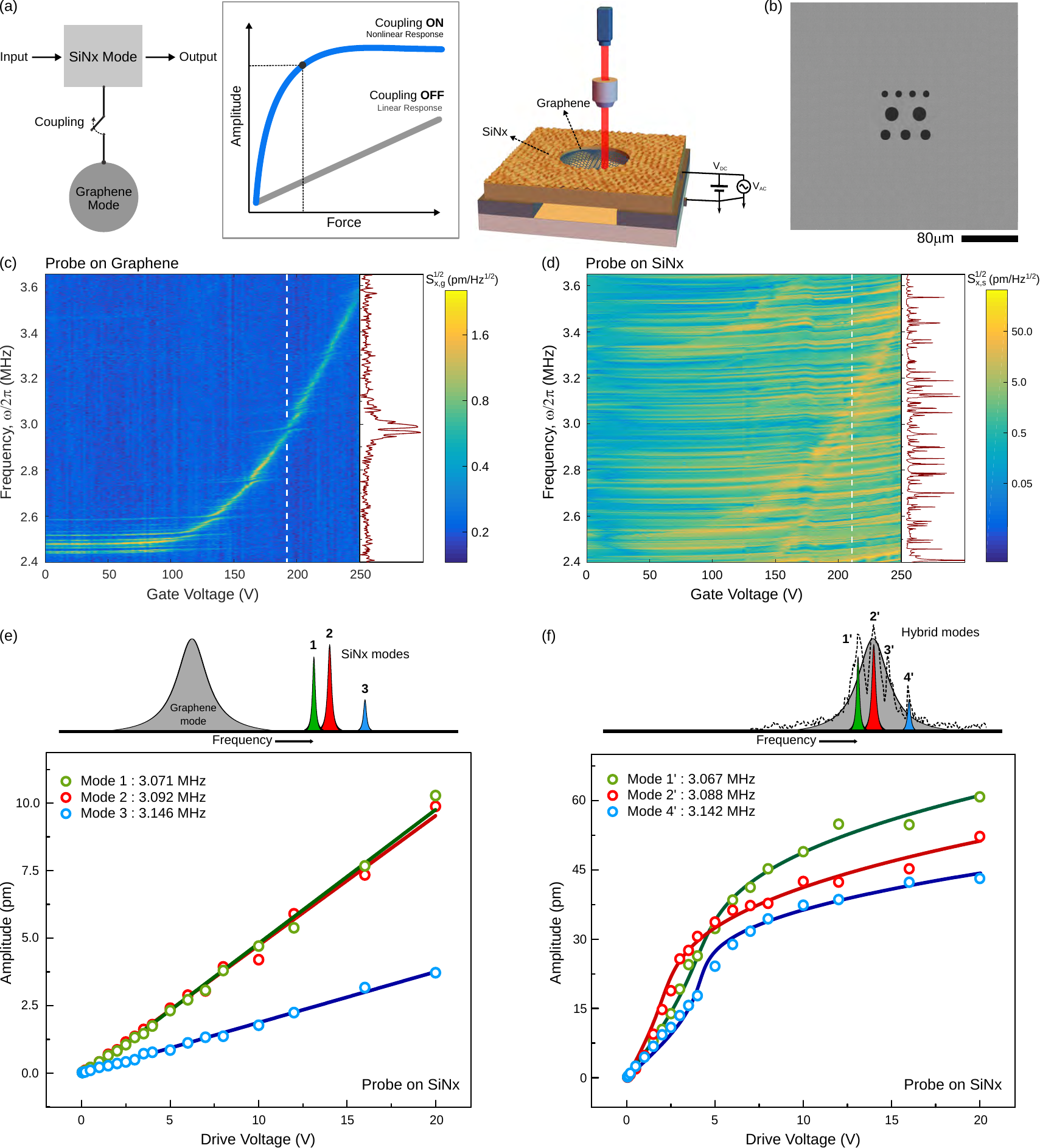}
	\caption{\textbf{Nonlinear response of graphene-SiNx modes:} \textbf{(a)} A cartoon depicting gate tunable, linear to nonlinear response of a SiNx resonator mode, due to its strong coupling to graphene. We observe oscillations of the hybrid mode with a confocal microscope, either focused on an atomically thin graphene or on the large area SiNx resonator surface (3d cartoon). \textbf{(b)} Scanning electron micrograph of the device with the large area SiNx resonator ($320\times 320\times 0.3\, \mu \rm m^3$, in grey) with graphene deposited onto $20, 15,$ and 10 $\mu \rm m$ diameter holes etched on it. The  $\rm 20\,\mu m$ diameter, intact graphene drum, coupling to SiNx is the focus of this study (see SI). \textbf{(c)} Thermally driven $20\, \rm \mu m$ graphene fundamental mode dispersion with the d.c. gate voltage interacting with array of SiNx modes. The right panel shows a cross-section at 194 V, where graphene hybridizes strongly with a single SiNx mode. \textbf{(d)} Corresponding dispersion of SiNx modes with dc gate voltage. The dispersion profile of fundamental modes of both the 20 $\rm \mu m$ and the 15  $\rm \mu m$ diameter graphene resonators can be observed, imprinted on SiNx dispersion, a signature of their backaction force. \textbf{(e)} Peak amplitude of three SiNx modes under direct a.c. drive shows linear response when not in resonance with the graphene mode. \textbf{(f)} When on resonance with low-$Q$ graphene mode, response of the SiNx modes modifies due the back action of graphene, becoming nonlinear (dots). Solutions for steady-state amplitude of a Duffing oscillator fit well with the data points (solid line) (see SI).} 
\end{figure*} 

\noindent

\section*{Nonlinear response of Graphene - Silicon Nitride Hybrid}

The device consists of a 300 nm thick SiNx resonator of dimensions 320 $\times$ 320 $\mu$m$^2$, with through holes of diameters $10,\,15,$ and $20$ $\mu$m etched on to it (fig. 1b) and monolayer (CVD) graphene is deposited on the holes~\cite{Nicholl-15}. Both graphene and SiNx resonators are actuated electrostatically with a highly doped silicon back gate, separated by an insulator which results in a net separation of 10 $\mu$m between graphene/SiNx and back gate. A fiber based confocal microscope, as part of a path stabilized Michelson interferometer, is used to detect optical signals reflected from the sample (3d cartoon of fig. 1a). 

When the microscope is focused on a 20~$\mu$m freely standing graphene membrane, we observe thermally driven modes, corresponding to that of a circular resonator with anisotropic tension~\cite{Singh-18,Davidovikj-16}. The modes are tunable in excess of 1 MHz with a d.c. gate voltage (fig. 1c) and as we tune from 0 V to 250 V, we see distinct avoided level crossing that signals hybridization with modes of a SiNx resonator (fig. 1c and right panel). In a recent work we have found that a bilinear coupling model fits well with observed hybridized Brownian spectra. From the fitting, we extract the quality factors: $Q_g\sim 254$ for graphene and $Q_s\sim 3800$ for SiNx~\cite{Singh-18}. The effective mass of the graphene mode, estimated from dispersion of fig. 1c is estimated at $m_g \sim 10m_0$, where $m_0$ is the mass of a single layer of graphene~\cite{Singh-18}. We estimate the effective mass, $m_s$, of SiNx from its dimensions and density, and find it to be $\sim 10^{4}m_g$~\cite{Singh-18}. 

Comparatively heavier mass of SiNx results in smaller amplitude for the Brownian power spectrum, which is below our detection sensitivity (see supporting information (SI)). However, when the device is actuated with a.c. gate voltage and the microscope is focused on SiNx surface (away from graphene), we observe dense distribution of SiNx resonator modes (fig.~1d). The mode-densities and their dispersion (fig. 1d, right panel) match well with simulated modes of a square membrane of comparable dimensions with an inbuilt tension of 80 MPa (see SI). 

When SiNx modes are not hybridized with graphene, the peak amplitudes increase linearly with applied a.c. gate voltage, up to a maximum amplitude of 20 V that we can apply in our experiment. In particular, we measure three modes of SiNx at frequencies 3.071~MHz (mode 1), 3.092~MHz (mode 2), and 3.146~MHz (mode 3) as shown in fig. 1e. When the low-$Q$ fundamental mode of graphene is tuned into resonance by applying a d.c. gate voltage $V_g = 210$ V, we observe frequency shifts of the three modes and a increased linear response due to back-action of the coupled graphene mode (see S.I.). Beyond a certain gate voltage, the response becomes nonlinear (fig.~1f). The data fits well with the model of ref.~\cite{Davidovikj-17}, for the steady state amplitude ($x_s$) of a forced oscillator with an additional nonlinear response that is cubic (Duffing) in $x_s$. From fitting, we extract the effective Duffing constant, $\beta^{s}_{\rm hyb}$, for example, for the hybrid mode $2^{\prime}$ to be $\beta^{s}_{{\rm hyb 2}} = 8.0(\pm0.8)\times 10^{21}$ $\rm N/m^3$ (see SI). This is one of the highest measured Duffing constants~\cite{Huang-16,Gajo-20}, orders of magnitude larger than graphene's reported values~\cite{Eichler-11,Davidovikj-17}. 

\noindent
What leads to such giant Duffing constants for the hybridized SiNx modes?
Aforementioned good fit of a Duffing model to the SiNx hybrid mode implies the source to be coupling graphene, a highly nonlinear Duffing oscillator. It is therefore critical to characterize the Duffing constant of the graphene resonator. For the graphene mode, we observe distinct signatures of Duffing-like hysteresis in response and asymmetric broadening due to nonlinear damping, when driven on resonance (see SI). Recent studies have characterized Duffing constant and nonlinear damping of driven graphene resonators by accurate fitting of data to theory~\cite{Eichler-11, Davidovikj-17}. However, in our device, hybridization of graphene with multiple SiNx modes results in a complex spectra (see SI). Instead, we develop an alternative methodology to estimate the nonlinear parameters. \\

\noindent
\section*{Mechanical frequency comb on Graphene surface}
The spectra simplifies, when we drive the system parametrically at a frequency that is sum of two dominant hybrid modes (fig.~2a)~\cite{Singh-18}. Parametric drive leads to gain in only two specific hybrid modes that are phase matched as opposed to the scenario of direct driving where multiple modes may interact. From the corresponding spectra, we develop a methodology to estimate Duffing constants of bare graphene as well as that of the hybrid modes as measured on graphene and on SiNx resonator surfaces.
\\

\begin{figure*}\includegraphics[scale=0.55]{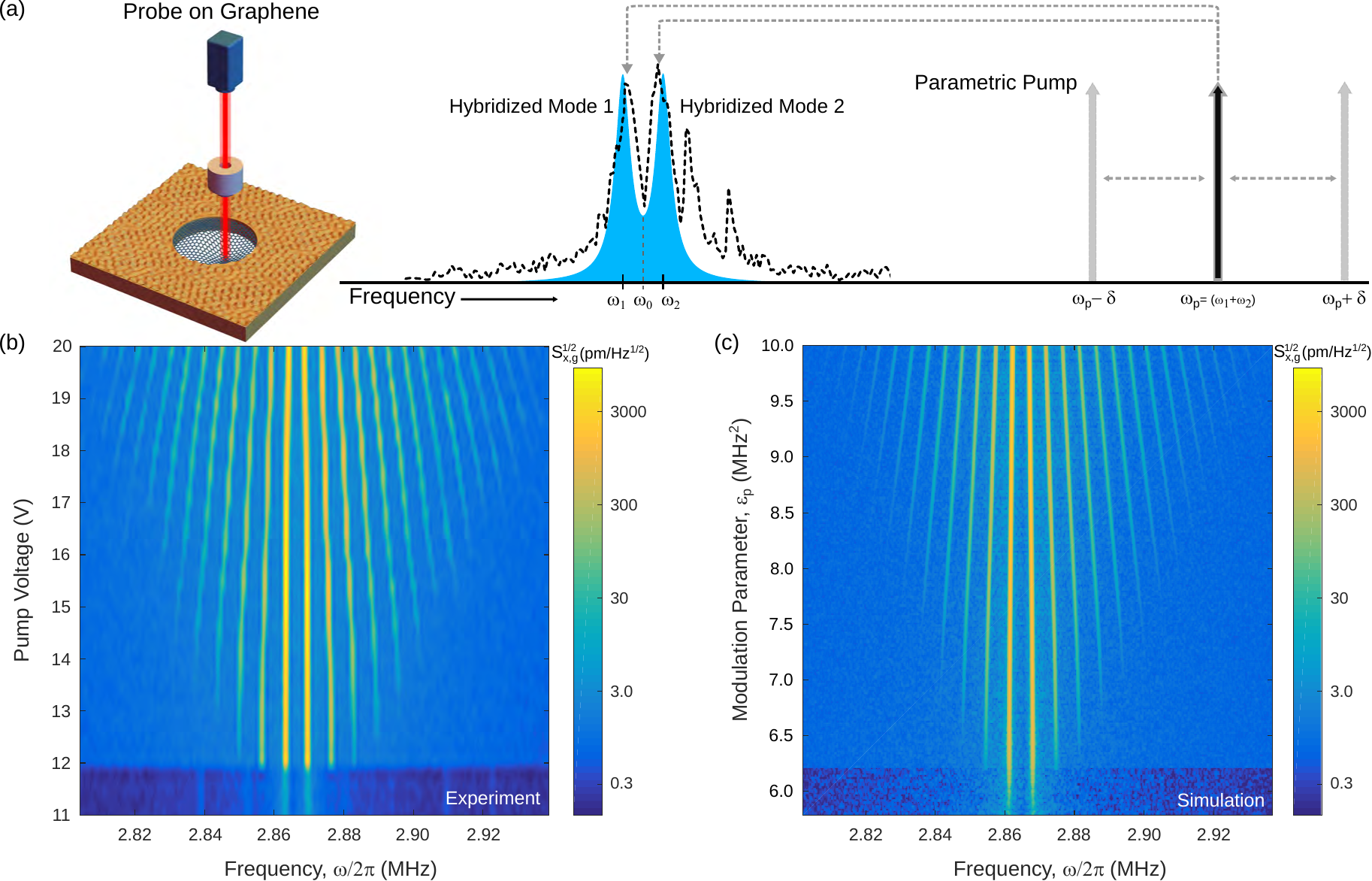}
	\caption{\textbf{Parametrically driven modes on graphene:} \textbf{(a)} Cartoon depicting graphene-SiNx hybrid mode, probed on graphene, under parametric pumping at the sum of frequencies of the hybrid modes. Detuning of the pump from this resonance is defined as $\delta$. \textbf{(b)} Observation of phase coherent frequency comb in the parametrically instability regime, preceded by non-degenerate parametric amplification. The plot shows displacement power spectrum \emph{measured on graphene}, with pump voltage scanned from 11~V to 20~V. \textbf{(c)} Corresponding numerical simulation of the coupled modes of linear SiNx and nonlinear graphene, driven with a parametric pump of strength $\epsilon_p$ (see SI). Calibrating the amplitudes to fig. 2b gives an estimate of the Duffing constant and the nonlinear damping of the graphene.} 
\end{figure*}

To parametrically drive the system, we first tune the fundamental mode of graphene at a frequency, $\omega_0$ = 2.865 MHz, at which it strongly hybridizes with a SiNx mode. With the microscope focused on the graphene, hybridization of modes is distinctly visible in the form of a splitting into two modes at frequencies $\omega_1$ and $\omega_2$ (say).  We simultaneously apply an a.c. gate voltage (parametric pump) at exactly twice the resonant frequency, $\omega_0$ (fig.~2a). As the amplitude of the parametric pump voltage is increased, up to a threshold voltage $V_c=$ 11.9 V we observe gain in both the hybridized modes (see SI and ref. ~\cite{Singh-18}). Above threshold, new frequency components on either side of the two hybridized modes develop. The number of such modes increases with increasing pump voltage, eventually spanning out into a ``comb" like pattern (fig.~2b)~\cite{ Ganesan-17,Ganesan-18}. We next develop a theoretical model to account for the new generated comb lines, towards estimating nonlinearities in the system.\\

\begin{figure*}\includegraphics[scale=0.053]{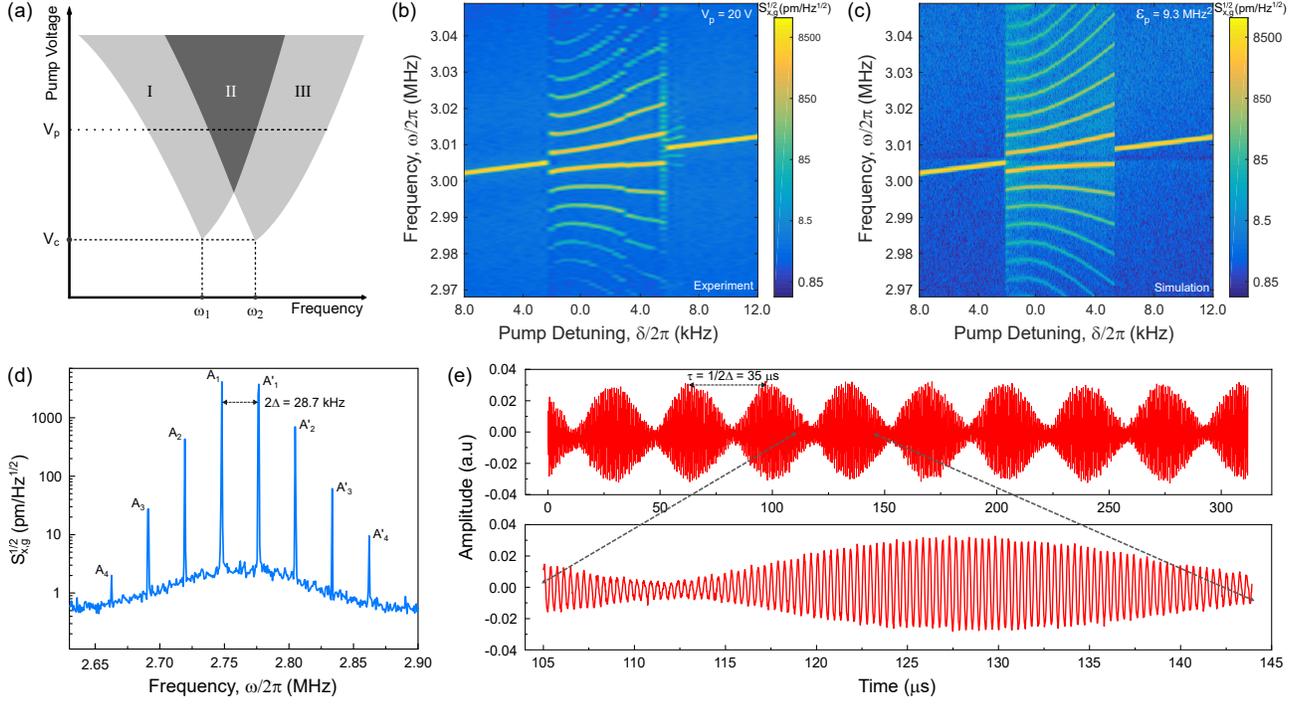}
	\caption{\textbf{Quantifying frequency comb:} \textbf{(a)} Cartoon of frequency response of a parametrically driven hybrid oscillator with two hybridized modes with frequencies $\omega_1$ and  $\omega_2$. Above a threshold, each mode develops a tongue shaped instability region. Accordingly one expects single frequency self-oscillation regimes (grey shades, I and III) while in the region of overlap, the system oscillates with two frequencies which mix to produce side bands (dark grey, region II). While the left edge of the envelope is due to the tongue of the hybridized mode $\omega_1$, the right edge is due to $\omega_2$. This insight leads to an estimation of the nonlinear coefficients of the hybrid modes.  \textbf{(b)} To test the hypothesis of fig. 3a, we scan the pump frequency ($\delta$) at a fixed a.c. pump voltage (black dashed line). We indeed observe a multi mode, comb spectra (region II) sandwiched between two single mode self-oscillation regimes (I and III). \textbf{(c)} Numerical simulation varying pump detuning matches well with the experimental observation. \textbf{(d)} Cross section of the multi-mode spectrum with mode separation of ${\sim28.7}\,{\rm kHz}$. From the ratios of amplitudes, we estimate nonlinear coefficients of the hybrid modes, \emph{measured on graphene}. \textbf{(e)} Proof of phase coherence: measured time trace of the generated signal of fig. 3d shows a pulse sequence that is Fourier-transform limited, confirming that the pulse is phase coherent.} 
\end{figure*} 
\section*{Theoretical Model} A model of graphene as a 1d oscillator with a quality factor $Q_g$, a third-order nonlinear response described by an effective Duffing constant ($\beta^g_{\rm bare}$) along with nonlinear damping ($\eta^g_{\rm bare}$) ~\cite{Eichler-11,Lifshitz-09,Nayfeh-07}, coupled to a linear SiNx resonator mode explains the observations well. Specifically, we simulate the following  set of equations:

 \begin{multline}
\ddot{x}_g = -\frac{\omega_g}{Q_g} \dot{x}_g -\frac{\eta^{g}_{\rm bare}}{m_g} {x_g^2} \dot{x}_g -\frac{\beta^g_{\rm bare}}{m_g} {x_g^3} \\
-[\omega_g^2 +\epsilon_p \cos(\omega_p t)] x_g  - \frac{\alpha}{m_g} x_s, 
\end{multline} 
and
\begin{equation}
\ddot{x}_s= -\frac{\omega_s}{Q_s} \dot{x}_s -\omega_s^2 x_s - \frac{\alpha}{m_s} x_g.
\end{equation} 
Here $\alpha$ is an effective coupling constant, $x_{g,s}$ are the amplitudes of vertical displacements of graphene ($g$) and SiNx ($s$) resonators modes respectively, and $\epsilon_p$ denotes the magnitude of the parametric drive. $\beta^g_{\rm bare}$ is the Duffing constant of the bare graphene resonator mode while $\eta^{g}_{\rm bare}$ is the coefficient of nonlinear damping.

Numerically simulated spectra is in agreement with observations (see fig. 2). In particular, we find linear coupling explains the frequency comb, as opposed to nonlinear couplings in earlier works~\cite{Karabalin-09,Mahboob-12,Cao-14,Mahboob-16,Ganesan-17,Seitner-17,Ganesan-18,Czaplewski-18}. By fitting simulated spectra to measured spectra, we get an estimate for the Duffing constant of the bare (non-hybridized) graphene mode~\cite{Storch-18} to be $\beta^g_{\rm bare} = 5.8\times10^{13}$ $\rm N/m^3$ with a nonlinear damping coefficient $\eta^g_{\rm bare} = 9.7 \times10^{6}$ $\rm Ns/m^3$, in close agreement with recent measurements~\cite{Davidovikj-17, Storch-18}.\\

\begin{figure*}\includegraphics[scale=.27]{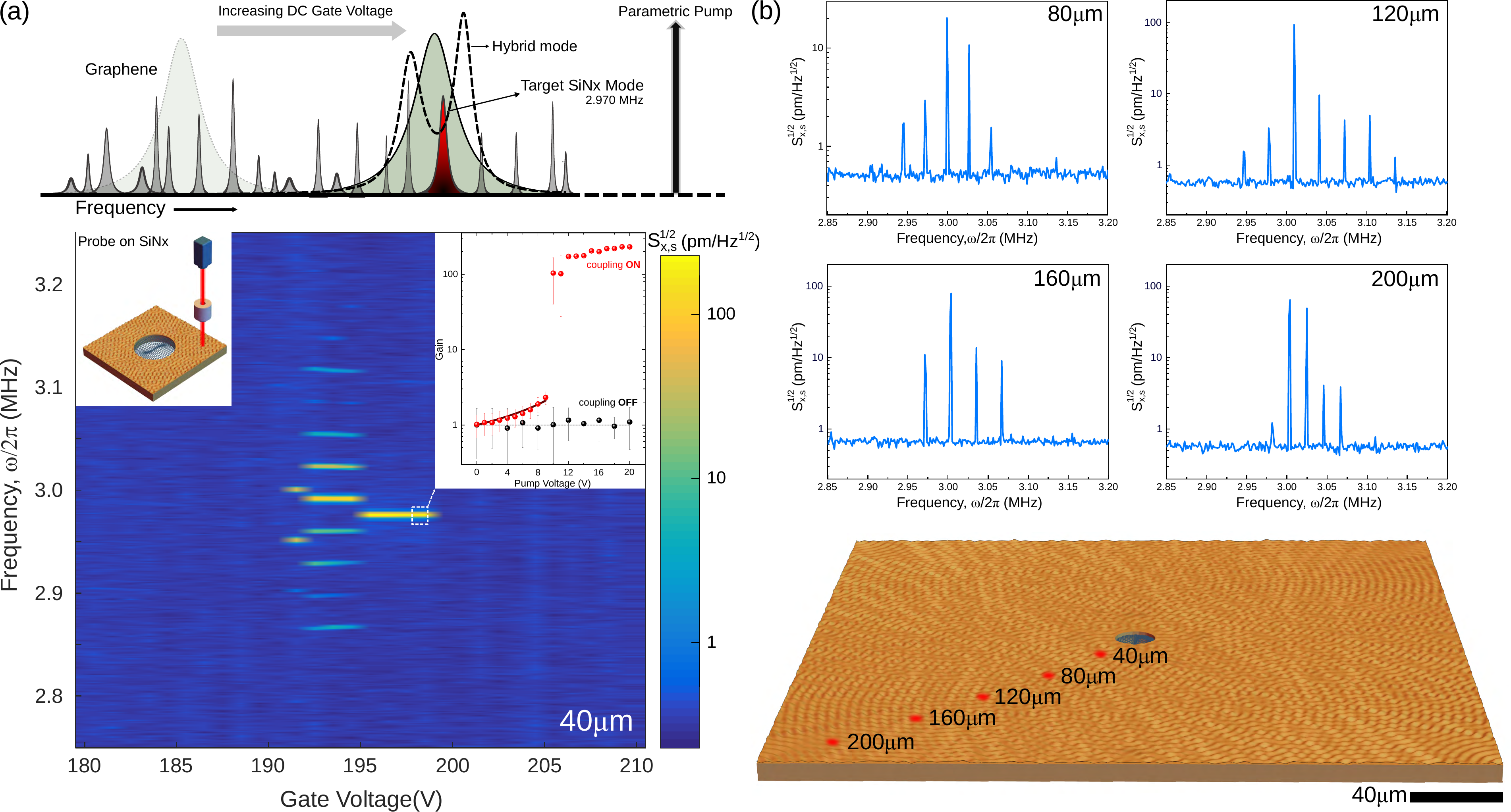}
	\caption{\textbf{Parametrically driven modes on SiNx:} \textbf{(a)} When probed on SiNx, the measured displacement power spectra shows no response to our maximum paramteric drive amplitude (inset). However, when a graphene mode is tuned into resonance, we observe parametric gain below a threshold (inset), along with generation of the frequency comb. We again use the comb lines to estimate nonlinear coefficients of the hybrid, as measured on SiNx resonator. \textbf{(b)} As we move the microscope away from graphene remaining focused on SiNx, we observe nonlinear response the hybrid modes leading to multimode spectrum at far distances (4 measurements shown as red spots on a device cartoon along with one at 40 $\mu$m on left, fig. 4a). Nonlinear response of the hybrid modes, for such small displacements of the SiNx surface, results in a giant Duffing constant $\beta^s_{{\rm hyb}}$ for SiNx. } 
\end{figure*}

The model further suggest that the essential physical mechanism behind the comb can be understood with the normal (hybrid) modes, even in the strong driving regime (fig. 3). In particular, it is well understood that response of a parametrically driven mode becomes unstable above a threshold: beyond threshold, the instability region extends to form a tongue shaped region~\cite{Hsu-63,Landau-82,Rand-12,Kovacic-18}. The envelope of the tongue is set by pump amplitude, nonlinear frequency, and damping. Therefore, for two hybridized modes with frequencies, $\omega_{1}$ and $\omega_{2}$, there should be two such independent instability tongues (fig. 3a). Consequently, there ought to be a region of overlap (dark region II, fig. 3a)~\cite{Hansen-85}. While region I and region III correspond to self-oscillation of hybrid modes 1 and 2 respectively, in the overlap region II the system is multi-periodic. Moreover, in the instability region, large amplitude leads to strong nonlinear response. One therefore expects mixing of two accessible frequencies ($\omega_{1}$ and $\omega_{2}$) in the overlap region II.  At a specific parametric drive amplitude (dotted line in fig. 3a), one thereby expects to observe these three regions.

We indeed observe these three regions when we vary the pump frequency across $\omega_p=2\omega_0$ at a fixed drive amplitude, $\rm V_p$ (dotted line in fig.~3a and also fig.~2a). In particular, the frequency is scanned over a range of 20 kHz around $2\omega_{0}$, keeping its amplitude fixed at $V_p$= 20 V (fig. 3b). We observe the single frequency self-oscillation regions I and III, on either side of region II that is characterized by the frequency comb (fig.~3b). Corresponding experimental observations match well with simulations (fig.~3c).

\section*{Estimating Graphene Nonlinearity}

Fig.~3a indicates that the right and the left boundaries of region II correspond to the instability tongues of the hybridized modes, viz., mode 1 and mode 2, respectively. One can then ascribe the observed asymmetry of the envelope of fig.~2b to differing effective nonlinearities of the two modes. Furthermore, from the experimentally measured amplitudes ($A_i$ and $A'_i$, $i$=1,2,$\cdots$) of the new comb lines generated due to cubic nonlinearity and nonlinear damping, we estimate the average nonlinear coefficients as $\beta^g_{\rm hyb 1}$= $ 1.3(\pm 0.4)\times 10^{15}$ $\rm N/m^3$, ${\beta^g_{\rm hyb 2}}$ = $8.6(\pm 4.6)\times10^{14}$ $\rm N/m^3$, ${\eta^g_{\rm hyb 1}}$ = $ 7.0(\pm 2.2)\times 10^{7}$ $\rm Ns/m^3$, and ${\eta^g_{\rm hyb 2}}$ = $ 4.8(\pm 2.5)\times 10^{7}$ $\rm Ns/m^3$, for the two hybrid modes 1 and 2, as measured on the graphene surface (see SI). The estimated parameters from data are in good agreement with numerical simulations (see SI), substantiating our methodology.

There is a phase relationship of the generated modes with respect to the fundamental modes, at frequencies $\omega_{1,2}$ , and in principle, the nonlinear coefficients can also be estimated by carefully measuring relative phases of the generated modes. For the spectrum of fig. 3d, we observe pulses in time domain (fig. 3e).  Repetition rate of the pulses correspond to inverse of 2$\Delta = \omega_1-\omega_2$, pulse width to inverse of the envelope of the generated comb while the carrier frequency to inverse of the carrier frequency $\omega_0 = 2.760$ MHz (fig. 3d). The frequency comb of the hybrid mode is therefore phase coherent and Fourier transform limited.

\section*{Induced frequency comb on Silicon Nitride surface}
It can be noted that due to widely varying masses and quality factors of the two physical resonators, values of Duffing constants of a hybrid mode would differ when measured on graphene or on SiNx resonator surface (see SI). Interestingly, signature of the comb spectrum is also observable, when the microscope is focused on the surface of the SiNx resonator (fig. 4). However, the amplitude of oscillations is orders of magnitude smaller than that on graphene, due to significantly heavier mass of SiNx. Accordingly, signatures of measured spectrum are less pronounced. Nevertheless, we use the developed methodology to estimate Duffing constants of the hybrid modes on SiNx. In fig. 4, we focus on a SiNx mode at 2.970 MHz, while applying a parametric drive at twice its resonant frequency. When the fundamental graphene mode is off-resonant, we do not observe any parametric gain (blue region on the left of fig. 4a and black dots in the right inset). However, when the graphene mode is tuned across resonance with a gate voltage between $193\,{\rm V}$ (approx.) to $200\,{\rm V}$, we observe generation of frequency comb as well as single frequency self-oscillation regime. Furthermore, the induced nonlinearity of the hybrid mode extends over the entire SiNx surface and we observe generation of combs at distances in excess of 200 $\mu$m from the edge of the graphene drum that is only 20 $\mu$m in diameter (fig.~4b, bottom panel). Essentially, the localized mode of the graphene acts as a defect center, on the large area oscillating mode of SiNx. From amplitudes of the generated modes, we estimate the effective Duffing constant and nonlinear damping of the hybrid modes to be  ${\beta^s_{\rm hyb 1}}$= $ 3.4(\pm 0.1)\times 10^{23}$ $\rm N/m^{3}$, ${\beta^s_{\rm hyb 2}}$= $ 6.3(\pm 5.3)\times 10^{22}$ $\rm N/m^{3}$ and ${\eta^s_{\rm hyb 1}}$= $ 1.8(\pm 0.1)\times 10^{16}$ $\rm Ns/m^{3}$, ${\eta^s_{\rm hyb 2}}$= $ 3.3(\pm 2.8)\times 10^{15}$ $\rm Ns/m^{3}$ respectively, as \emph{measured on SiNx surface} (see SI). The estimates are in close agreement with effective Duffing constant ${\beta^{s}_{{\rm hyb}}}$ measured on SiNx in fig. 1f.\\
\noindent
\section*{Discussion}
To conclude, here we have explored nonlinear response of graphene-SiNx hybrid modes and developed a methodology to quantify corresponding nonlinear coefficients, as measured on graphene and SiNx resonator surfaces. The observations suggest that the coupled system can be described by two uncoupled, nonlinear hybrid modes. Measured Duffing constants of these hybrid modes on SiNx surface are found to be in excess of eight order of magnitude larger than that on graphene. This indicates that nonlinear response is highly efficient on SiNx surface, setting in at displacement scale that is two orders of magnitude smaller, at 30.4 pm, compared to measurement on graphene.

It is remarkable that an atomically thin resonator generates a significant backaction force ($F^{\rm ba} = \alpha x_g$) on SiNx. Based on the $F^{\rm ba}$, a perturbative estimate yields $\beta^s_{\rm hyb} \propto \alpha\beta^g_{\rm bare}m_s^3/m_g^4$ (see SI) and indicates graphene to be a powerful candidate to induce such giant nonlinearity due to three primary factors: firstly, pristine graphene robustly couples to SiNx substrate via stable electrostatic forces resulting in a large coupling strength ($\alpha$). This also leads to better device yield. Secondly, low mass of graphene ($m_g$) results in a large amplitude  ($x_g$) of  oscillation, boosting the force further. Finally, exceptionally large Young's modulus results in large nonlinear response ($\beta^{g}_{\rm bare}$)~\cite{Davidovikj-17} to an applied force. Large gate tunable backaction force of graphene thereby emerges as the dominant mechanism behind observations in this work.

For our device, the tension of SiNx resonator is merely 80 MPa~\cite{Fink-16}, leading to comparatively lower quality factors ($\sim 3000$ on average), along with a dense distribution of SiNx modes (fig. 1d). An immediate improvement can therefore be towards increasing inbuilt tension of the SiNx resonator, so that one can resolve mode shapes distinctly~\cite{Yang-19} and observe graphene induced interaction between SiNx modes of quality factors in excess of $10^6$, possibly in a quantum regime. With such improvements, the hybrid device proposed here can provide a powerful platform for generating mechanical squeezed states in precision measurements and controlled interactions of mechanical modes both in classical and quantum domains at room temperature~\cite{Reinhardt-16,Norte-16,Patil-15}.

\section*{Materials and Methods }

\subsection*{ Sample preparation:}
Silicon Nitride membranes (thickness 300 nm) are fabricated by depositing low-stress silicon-rich silicon nitride on both sides of a silicon chip. An array of holes of 10, 15 and 20 um diameter is then patterned in the nitride using standard fabrication procedures. A metallic contact (20 nm Au) is deposited onto the top surface of the SiNx to facilitate electrical gating. Monolayer chemical vapor deposition (CVD) graphene with flake size of $\sim 90\times110 ~\rm\mu m^2$  is then transferred onto holes in the nitride membranes. We use a high-quality atmospheric CVD growth and wet transfer. The samples are subsequently annealed in an $\rm Ar-H_2$ environment at $350^\circ C$. The graphene membranes remained clamped to the sample chip via Van der Waals interactions forming suspended circular graphene membranes.

\subsection*{Experimental setup:}
We use a fiber based confocal microscope (see SI) with a spot size of 4 $\rm \mu m$ to optically probe our graphene-SiNx hybrid device. The microscope forms one arm of a Michelson interferometer while the reference arm is actively stabilized against drifts or fluctuations through a feedback form PI lock. A frequency and amplitude stabilized external cavity diode laser (ECDL) ($\rm \lambda$ = 780 nm) is used as an optical probe. All the measurements were conducted at probe power of $\sim$400 $\mu$W. The sample is placed inside a vacuum chamber ($10^{-2}$ $\rm mbar$) with high voltage electrical leads for gate control. The entire chamber assembly is mounted on a 3D scanning stage with active position locking. For detection, we use a balanced photo-detector with a detection bandwidth of 45 MHz. We position the sample by actively monitoring the generated 2-D confocal image, which helps in selecting the relative probe position and to lock the microscope there. The photo-current signal is analyzed with spectrum analyzer and dual-lock-in-amplifier.

\section*{Acknowledgements}

We thank Srivatsan Chakram, Deb Shankar Ray, Edgar Knobloch, Siddharth Tallur, Mandar Deshmukh and Amit Agarwal for insightful discussions and comments.
We also thank Om Prakash for his numerous help
in construction of the experimental setup. A.S. acknowledges CSIR and K.B. acknowledges ERC grant no. 639739 and DFG TRR 227 for financial support. This work was supported under DST grant no. SERB/PHY/2015404.

\setcounter{equation}{0}
\renewcommand{\theequation}{S.\arabic{equation}}
\renewcommand\thefigure{S.\arabic{figure}}    
\setcounter{figure}{0}

\begin{widetext}
	\newpage



%
%
%
%
%
%
%





	\begin{center}
		\textbf{\large Supporting Information:\\ 
			Giant Tunable Mechanical Nonlinearity in Graphene-Silicon Nitride Hybrid Resonator}
	\end{center}
	



	%

\section{S.1. Experimental method}

\noindent
\textbf{Experimental Setup:} Fig. S.1 below illustrates the experimental setup. 
Fig. 1c, 2b, 3b, 3d, 4a and 4b in the main text and S.2-4, S.12 and S.15 in the supplement are taken with spectrum analyzer while Fig. 1d-f in the main text and S.5-6 and S.7 in the supplement are acquired by scanning the drive frequency from a lock-in-amplifier.\\
\begin{figure}[H]
	\centering
	\includegraphics[scale=0.45]{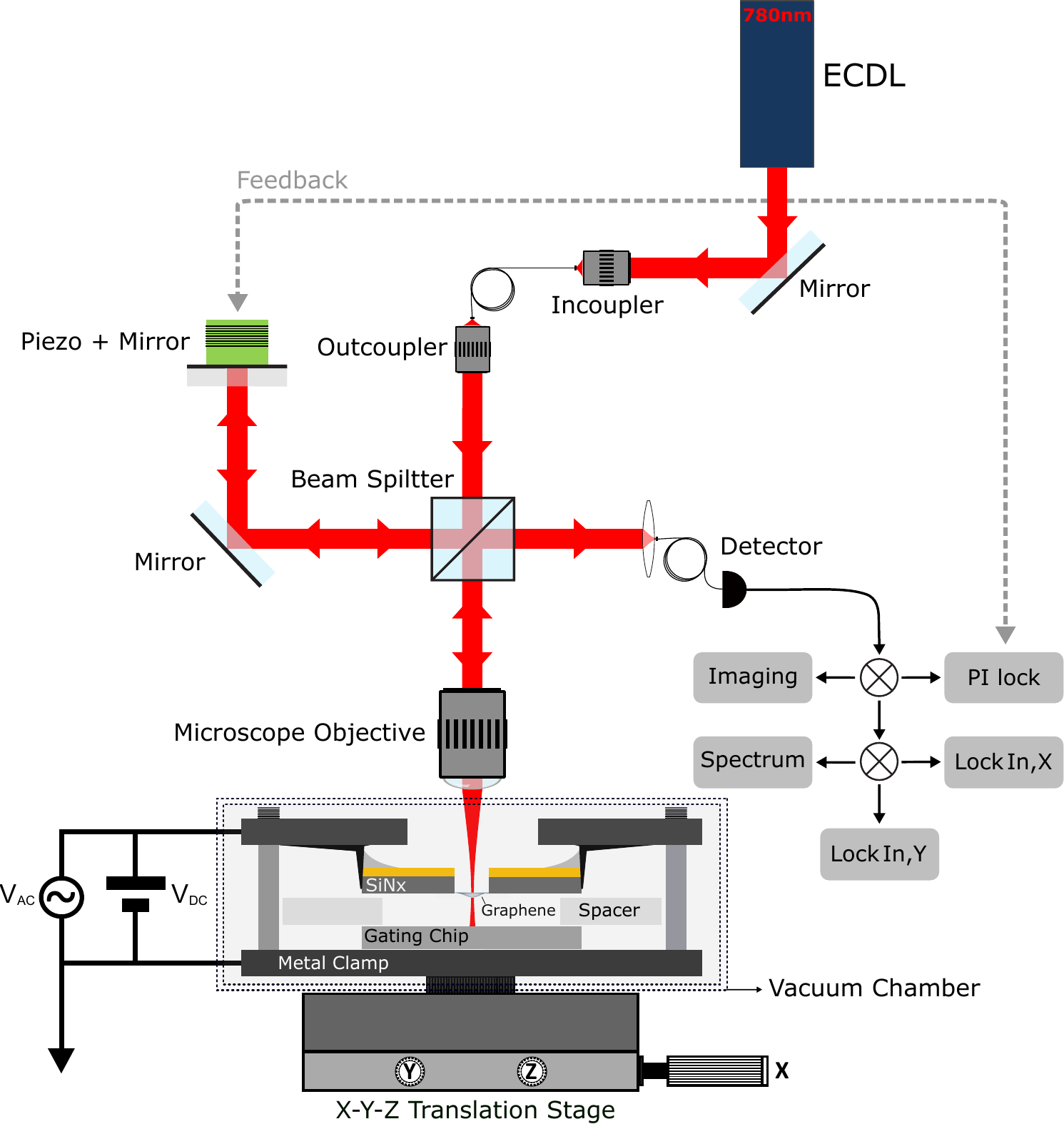}
	\caption{\textbf{Experimental Setup:} The figure illustrate the basic structure of the fiber based confocal microscope in interferometric arrangement used for the measurements.} 
\end{figure} 
\noindent

\pagebreak
\section{S.2. Calibration}
\subsection{A. Displacement calibration from hybrid Brownian spectrum}
\noindent
Displacements are calibrated by fitting the Brownian spectrum to a model that is based on coupled modes of graphene and SiNx resonators, denoted by displacements $x_g$ and $x_s$, respectively. For thermally driven graphene and SiNx modes we ignore the nonlinear terms. The equations of motion are then given by:
\begin{subequations}\label{first: a}

	\begin{equation}
	\ddot{x}_g+ \gamma^g_{bare}\dot{x}_g+ {\omega_g}^2 x_g - \frac{\alpha}{m_g} x_s= \frac{F^{th}_{g}}{m_g} 
	\end{equation}
	
	\begin{equation}
	\ddot{x}_s + \gamma^s_{bare}\dot{x}_s+ {\omega_s}^2 x_s - \frac{\alpha}{m_s} x_g=\frac{F^{th}_{s}}{m_s},  
	\end{equation}
\end{subequations}
where $\gamma^k_{\rm bare}$, $\omega_k $ and  $F^{th}_{k}$ $(k=g,s)$ represent linear damping, normal mode frequency and thermal forces acting on graphene and $\rm SiNx$ modes respectively. Coupling of graphene and SiNx modes is modeled by an effective interaction Hamiltonian, $H_{\rm int} = \alpha x_g x_s$\footnotemark \footnotetext{The bi-linear form of interaction is an approximate expression, derived from an effective Hamiltonian that can be expressed as:  $H_{\rm int} =-\frac{\alpha}{2}(x_g- x_s)^2 $, for small displacements of SiNx and with renormalized resonant frequencies.}. Solving the above coupled equations in Fourier space, the displacement power spectrum for the graphene resonator is:
\begin{equation}
S_{x,g}^{1/2}=\Bigg[\kappa\bigg(\frac{\frac{S_{F,g}^{th}}{m_g^2}\{(\omega_{s}^2-\omega^2)^2+(\gamma^s_{bare})^2\omega^2\}+\{\frac{S_{F,s}^{th}\alpha^2}{m_s^2m_g^2}\}}{[(\omega_{g}^2-\omega^2)(\omega_{s}^2-\omega^2)-\gamma^g_{bare}\gamma^s_{bare}\omega^2-\frac{\alpha^2}{m_g m_s}]^{2}+[(\omega_{g}^2-\omega^2)\gamma_{s}\omega+(\omega_{s}^2-\omega^2)\gamma^g_{bare}\omega]^{2}}\bigg)+S_{noise}\Bigg ]^{1/2},
\label{eqn:3}
\end{equation} 

\noindent

where $S_{F,k}= 4k_B T\gamma^k_{bare},(k=g,s)$ is the thermal force acting on graphene ($g$) and SiNx ($s$). The calibration factor, $\kappa$ along with all other free parameters are extracted by fitting experimental data, $ S_{v,g}^{1/2}$ $\rm (V/\sqrt{Hz})$ to the above equation. Using the calibration factor, the recorded spectrum is then converted into displacement spectrum.  
\\\\

\begin{figure}[H]
	\centering
	\includegraphics[scale=0.7]{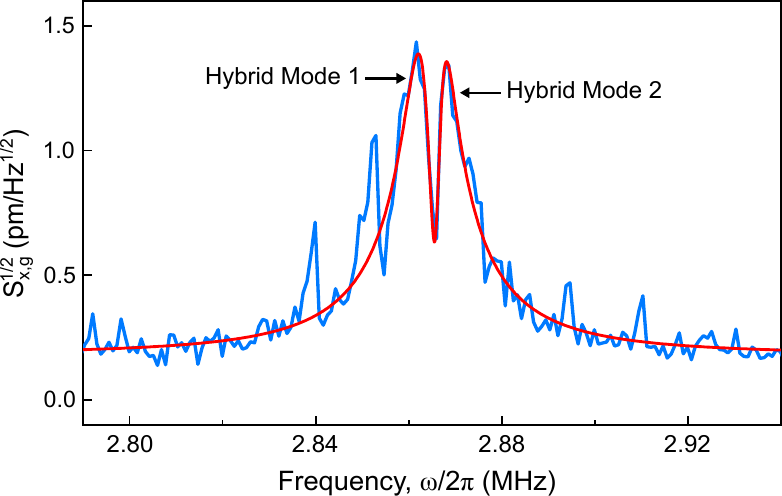}
	\caption{\textbf{Brownian spectrum:} The displacement spectrum of thermally driven mode of graphene-SiNx hybrid fitted with equation S2.} 
\end{figure} 
Extracted values of the fitting parameters for the Brownian mode corresponding to Fig. S.2 are listed below.

\noindent
$\kappa=1.921(\pm .074)\times10^{10}$ $\rm V^2/m^2$, $\omega_g/2\pi=2.8646(\pm .0002)\times 10^6$ Hz, $\omega_s/2\pi=2.8656(\pm .0001)\times 10^6$ Hz, $\gamma^g_{bare}/2\pi$=11.237($\pm$ 0.690)$\times10^3$ Hz, $ \gamma^s_{bare}/2\pi=0.744 (\pm 0.234)\times 10^{3} $ Hz, $\alpha/4\pi^2=1.978(\pm 0.095)\times 10^{-3}$ $\rm kgHz^2$.\\\\
This model can be extended for graphene interaction with multiple SiNx modes. The supplemental information of Singh, R. et.al.~\cite{Singh18} can be referred for more information.

We calibrate the amplitude of graphene mode by fitting its thermal or Brownian spectrum to a model of coupled 1d oscillators as shown above. The calibrated thermal mode is then used to calibrate the displacement spectrum of the parametrically driven mode. Similarly the thermal mode of SiNx is calibrated, either by fitting it to its Brownian spectrum, or using the calibration from graphene, where we carefully maintain all other microscope parameters.
 
\subsection{B. Mass estimation}
In order to estimate the mass of graphene, we fitted the fundamental mode dispersion of the graphene drum using continuum mechanics model~\cite{Chen-13}. We obtained a good fit for $m_g= 9.982 (\pm 0.008)$ $\times$ $m_0$ (where $m_0 = 0.625\times10^{-16}$ kg is the mass of pristine residue-free graphene resonator) and tension is $6.91(\pm0.01) \times 10^{-5}$ N/m (Fig. S.3). 

\begin{figure}[H]
	\centering
	\includegraphics[scale=0.75]{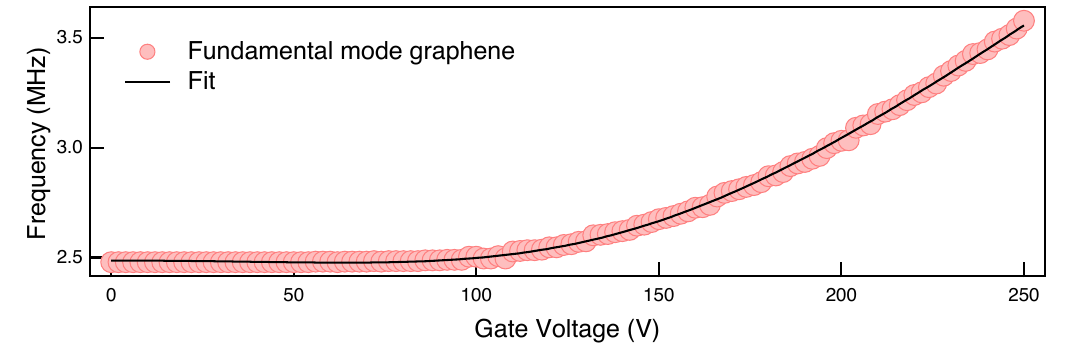}
	\caption{\textbf{Mass estimation:} Gate voltage dependence of the fundamental model for (red points) along with the fit (black curve) using the model in ref.~\cite{Chen-13}.} 
\end{figure} 

\subsection{C. Calibration with probe power}
We studied dependence of thermal motion of hybrid mode of graphene fundamental mode with increasing incident probe power. The spectrum at different probe powers were recorded and fitted with equation S2. The extracted calibration factor, $\kappa$ depicts linear scaling with the probe power. Invoking the equipartition theorem, we estimated the mode temperature and have not observed significant variation. Overall, these measurements suggest a range up to 1 mW of probe power over which the  graphene mode remains unperturbed. For all measurements reported in this work, we maintained the probe power at $\sim400$ $\rm \mu W$. 

\begin{figure}[H]
	\centering
	\includegraphics[scale=0.55]{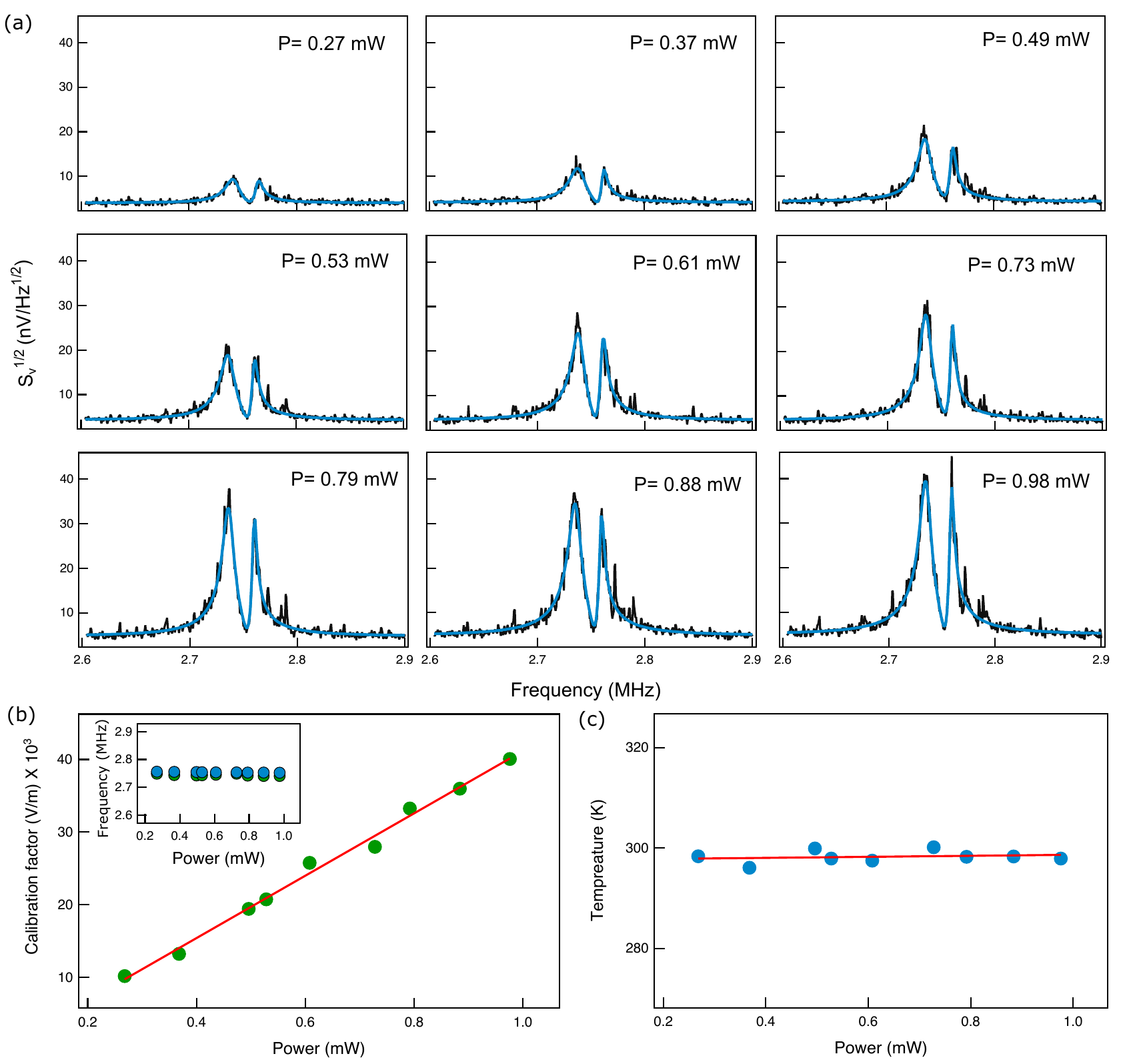}
	\caption{\textbf{Response of graphene-SiNx hybrid modes with increasing probe power:} \textbf{(a)} Voltage spectrum of a hybrid mode, measured on graphene and the data is fitted to a model of coupled harmonic modes (equation S2). Spectrums are shown for different probe powers. \textbf{(b)} Extracted displacement calibration factor (V/m), obtained from data fitting shows a linear scaling with increasing probe power. (Inset) modes frequencies of graphene (blue dots) and SiNx (green dots) resonators, showing a small but distinct frequency shift with increasing probe power. We do not fully understand the nature of the shift. Though this shift suggests decreasing tension of graphene resonator with increasing probe power. \textbf{(c)} The extracted temperature of the mode shows little variation with probe power.} 
\end{figure} 


\section{S.3. Graphene and SiNx resonator modes}

\subsection{A. Nonlinear modes of Graphene resonator}

\noindent
The sample consists of a large area SiNx resonator ($320\times320\times0.3$ $\rm \mu m^3$), with 20 $\mu$m, 15 $\mu$m and 10 $\rm \mu$m diameter circular holes etched on to it over which monolayer graphene is deposited, thereby forming suspended drums. The graphene gets clamped to SiNx at the edges via Van der Waals forces while rest of the part above the hole remains freely hanging.
When the microscope is focused on graphene, we observe its thermo-mechanical spectrum from the detected signal in a electronic spectrum analyser. 
\begin{figure}\includegraphics[scale=.57]{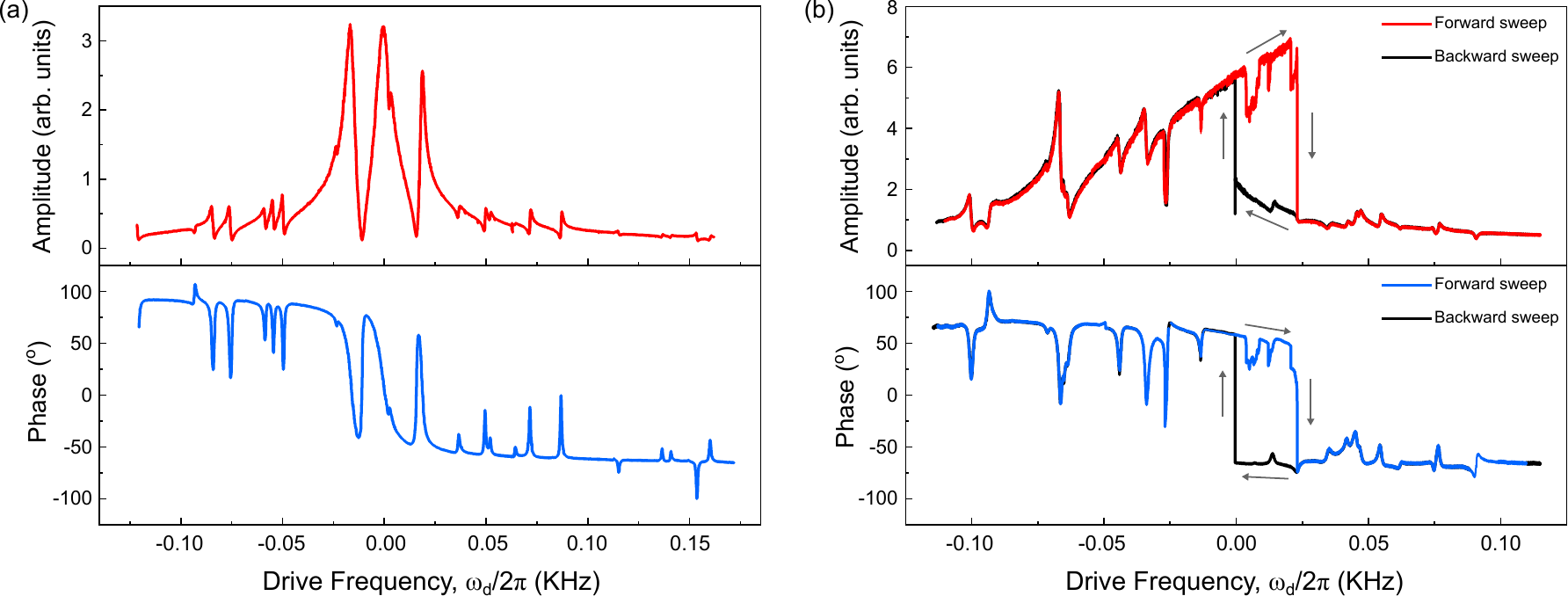}
	\caption{\textbf{Driven hybridized graphene modes:} \textbf{(a)} Amplitude plot of weakly driven (linear regime) graphene fundamental mode that is interacting with multiple SiNx modes. The bottom panel depicts the corresponding phase response. \textbf{(b)} At larger drive, amplitude and phase response exhibits asymmetric profile with hysteresis, a quintessential signature of cubic nonlinearity.} 
\end{figure}

\begin{figure}[H]
	\centering
	\includegraphics[scale=0.55]{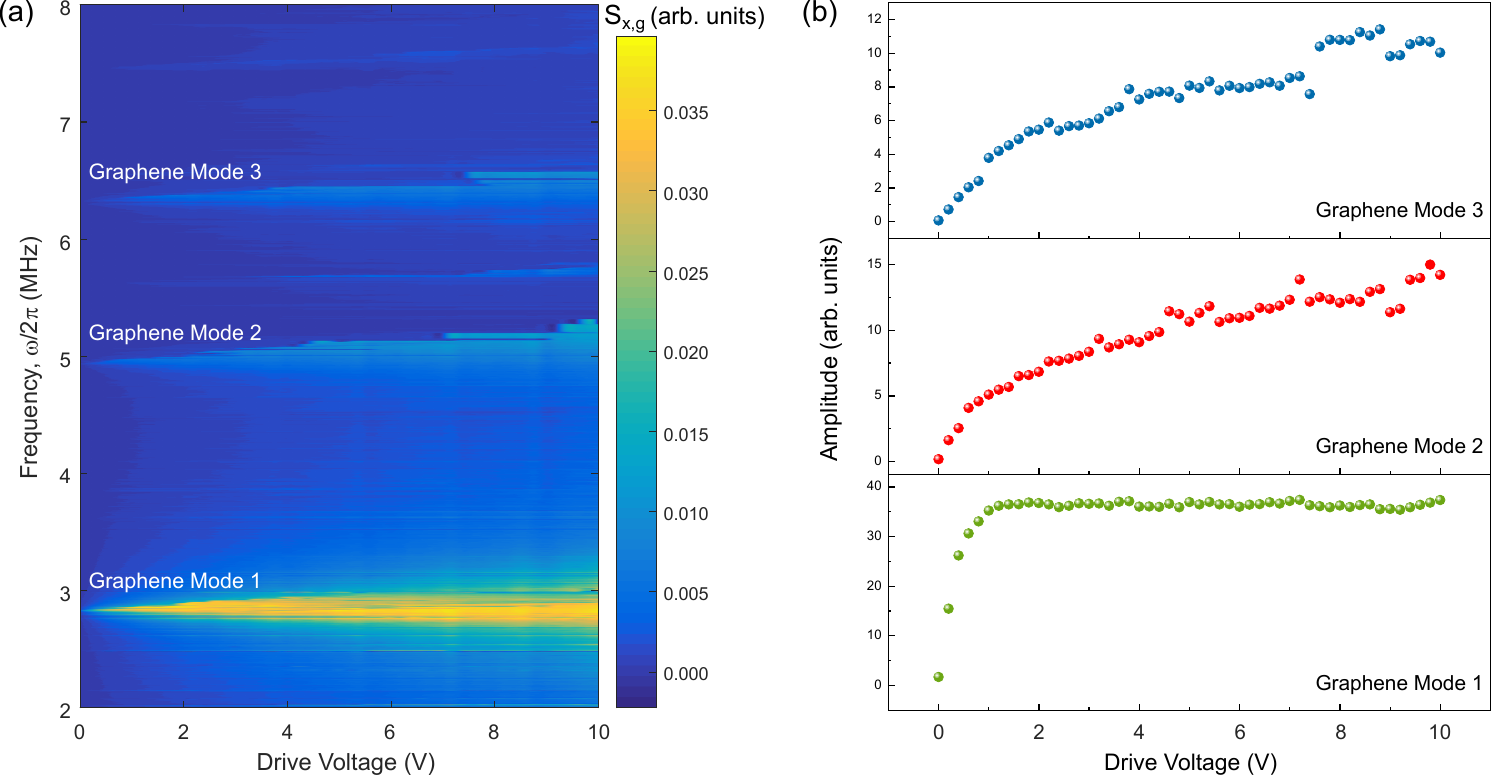}
	\caption{\textbf{Nonlinearity in graphene modes:} \textbf{(a)} Dispersion of driven graphene modes as a function drive voltage. \textbf{(b)} The peak amplitude of graphene modes as a function of drive voltage.} 
\end{figure}


The dispersion of fundamental mode of graphene as a function dc gate voltage (Fig. 1c) represents deviation of mode shape from usual Lorentzian shape, due to interaction of the graphene vibrational mode with densely packed, multiple SiNx modes. Such interactions lead to sharp dips and Fano-like asymmetry in graphene spectrum. The asymmetry gets more pronounced when the graphene mode is driven on resonance (Fig. S.5a). The corresponding phase profile shows an overall envelop corresponding to $\pi$-phase jump, as one crosses the broad graphene resonance. However, finer features in phase profile with sharp, intermediate $\pi$ phase jumps correspond to individual SiNx modes which are coupled to  graphene mode.  It can be noted that in general, each of these narrow SiNx modes have a unique coupling strength to the same graphene.

When driven harder, the broad graphene mode shows an asymmetric signature in spectra that is typical of a Duffing oscillator with cubic nonlinear response in displacement (Fig. S.5b). Forward and backward sweeps of drive frequency shows hysteresis in both the amplitude and phase.

\noindent
Fig. S.6a shows transition from a linear response to a Duffing-like nonlinear response for the graphene mode, coupled to multiple SiNx modes. The spectrum shows an increase in the FWHM (full width at half maximum)  with increasing drive voltage, pointing towards the existence of nonlinear damping. The amplitude of the  modes with drive voltage show saturation after a certain drive voltage and the critical voltage for saturation is specific to a given mode (Fig. S.6b).
\subsection{B. Modes of large area Silicon Nitride resonator}
\noindent
Silicon Nitride is a large area ($320\times320\times0.3$ $\rm \mu m^3$) resonator with through holes. Fig. S.7 shows the amplitude and phase of weakly driven SiNx modes. The modes are densely packed with quality factor in the range of 1000-4000. From COMSOL simulation, we estimate the inbuilt tension, $T_s$ $\sim$ 80 MPa.  

\begin{figure}
	\centering
	\includegraphics[scale=0.65]{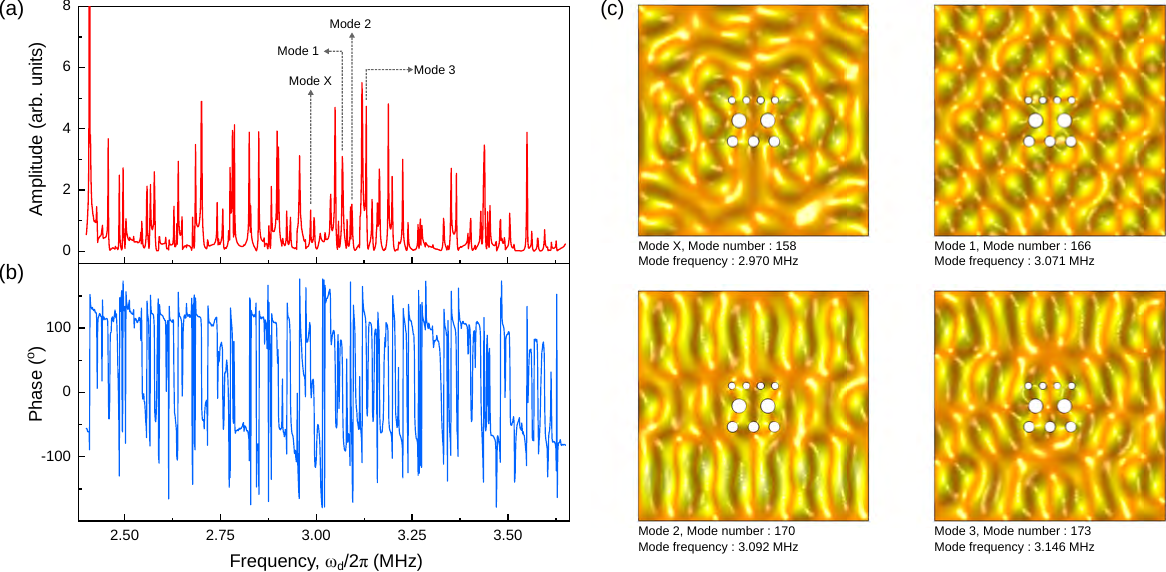}
	\caption{\textbf{Silicon Nitride Modes:} \textbf{(a)} Amplitude response of driven SiNx modes. In particular the modes marked as 1, 2 and 3 and their interaction with the fundamental graphene mode were studied in detail. \textbf{(b)} Corresponding phase profile of the SiNx modes. \textbf{(c)} Simulated spatial profiles of modes X, 1, 2 and 3 indicated with their mode number and frequency respectively. Simulations were performed using COMSOL.} 
\end{figure} 

\subsection{C. Linear response of the hybrid Silicon Nitride mode}

In Fig. 1e,f in main text, one can see a significant difference between uncoupled and hybridized SiNx mode response even in the linear response regime. This linear response behavior can be appreciated by noting (Fig. S8) that SiNx when coupled to graphene experiences two different forces: a direct capacitive force and a back action force from graphene.
Even in the linear regime, the backaction force experienced by SiNx from graphene is larger than the capacitive force under direct a.c. drive. This is the reason behind the larger displacement or steeper slope in linear regime of SiNx when on resonant with the graphene. 
To validate our assertion, we have conducted further experiments, comparing the two differing linear regimes. In particular, we consider the case when graphene is off resonant to the SiNx mode. We record the peak amplitude of vibration for SiNx and graphene modes independently, corresponding to the a.c. voltage. The Fig. S8b and c show the response of SiNx and graphene resonator modes, with the a.c. drive voltage, respectively. From fitting, we find the slope of graphene response (in linear region) is ~470 times larger than that of SiNx i.e, $x_g^c\sim470\times x_s^c$.

When the SiNx mode is on resonant to the graphene mode, an additional coupling dependent backaction force acts on the SiNx which can be written as follow:
$F_{ba}^{g\rightarrow s}=\alpha (x_g^c-x_s^c) \sim \alpha_g^c ,  x_s^c<x_g^c$.
The displacement experienced by the SiNx due to this backaction force is:
$x_s^{ba}=F_{ba}^{g\rightarrow s}\frac {Q_s}{m_s \omega_0^2}= \alpha x_g^c  \frac{ Q_s}{m_s \omega_0^2}$.

However, we have $x_g^c=470\times x_s^c$. Plugging this relation and typical values of other parameters ($Q_s=3000,\omega_0=2.86 ~\rm MHz, m_s=2.32\times 10^{-11} ~kg, \alpha=1.19\times 10^{-3}  ~\rm kgHz^2)$ in above equation, yields 
$\frac{x_s^{ba}}{x_s^c }=14$.
The result suggests that the displacement of SiNx mode due to backaction force is 14 times larger than that of capacitive force. This is consistent with the observation of Fig. 1e and f, main text, where a larger linear response is observed on SiNx, when on resonant with graphene.
It can be noted that such displacements are still smaller than the nonlinear threshold displacement of SiNx. However, when the graphene is driven into nonlinear regime, the response of the hybrid mode, as measured on SiNx also becomes nonlinear.

\begin{figure}
	\centering
	\includegraphics[scale=0.3]{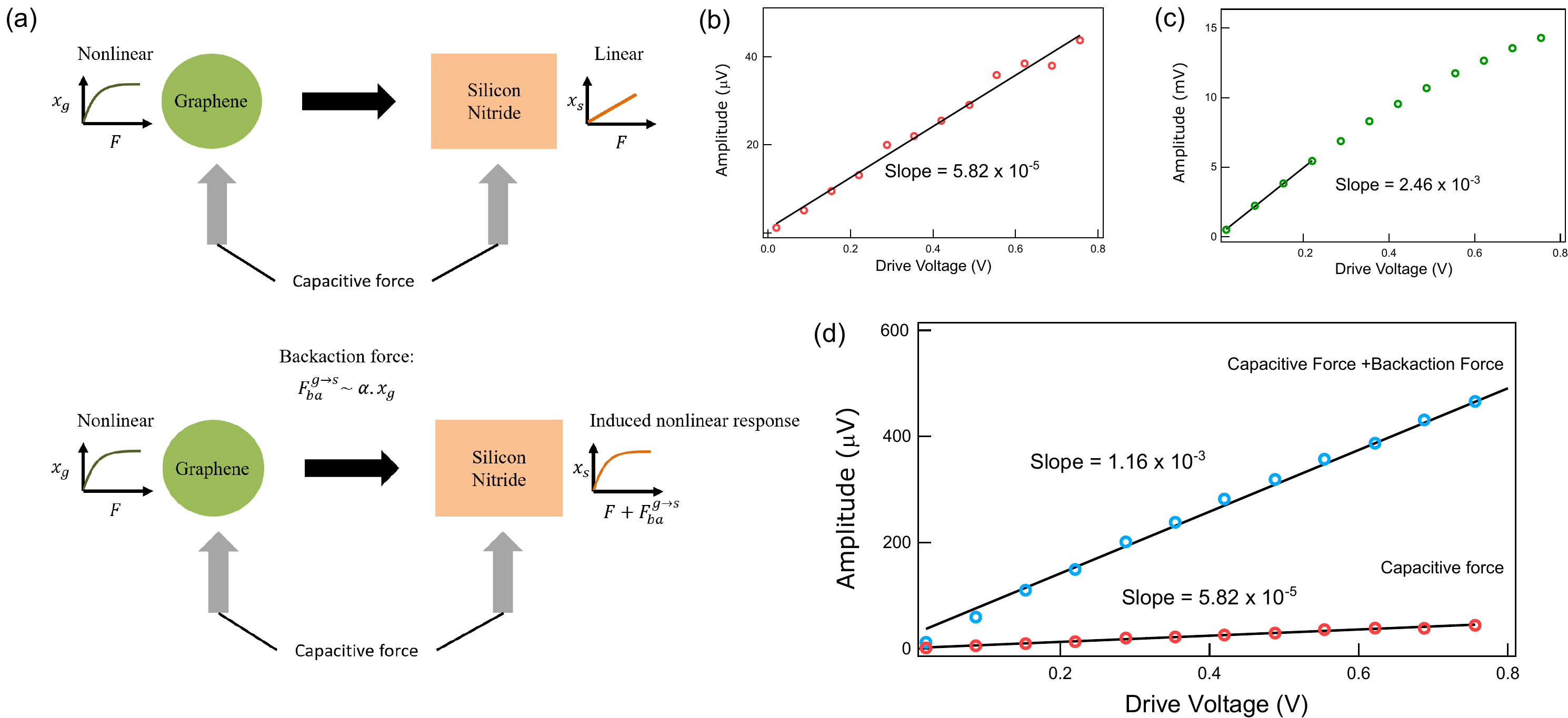}
	\caption{\textbf{Silicon Nitride response under different condition:} \textbf{(a, top panel)} Graphene far off resonant to SiNx mode i.e. coupling coefficient is zero. A capacitive force act on these resonators due to a.c. drives voltage. \textbf{(a, bottom panel)} When graphene is on resonant with the SiNx mode, the latter experiences a coupling dependent backaction force in addition to the capacitive force. The backaction force which itself is “nonlinear” coax a “linear” SiNx mode to display nonlinear behavior. \textbf{(b)} Amplitude response of SiNx with  drive voltage. \textbf{(c)} Amplitude response of graphene with drive voltage. \textbf{(d)} Amplitude vs. drive voltage of SiNx when on (blue) resonant and off (red) resonant to graphene.} 
\end{figure}

\subsection{D. Nonlinear response of hybrid Silicon Nitride modes under direct driving}
\noindent
We observe SiNx modes, respond linearly to external drive when it is off resonant from graphene mode. However, when we couple a graphene mode to the SiNx mode, its response becomes nonlinear with applied forces and shows saturation in amplitude above a critical displacement.

To quantify nonlinear response of the SiNx modes, we follow the procedure described in D. Davidovikj et. al.~\cite{David-17}. We first extract the \textit{slope} from the linear region of $x_s$ vs $V_{ac}$ plot (Fig. 1e, main text).
The rescaled force, $F$ corresponding to the $V_{ac}$ is given by,
\begin{equation}
F=slope\frac{\omega_s^2m_s}{Q_s}V_{ac}.
\label{eqn:4}
\end{equation} 
This rescaled force is plotted with $x_{s} $, the steady-state response of a Duffing oscillator, such that~\cite{David-17}:
\begin{equation}
\zeta F=(Ax_s^2+Bx_s^4+Cx_s^6)^{1/2}
\label{eqn:5}
\end{equation} 
where $C=\frac{9}{16}(\beta^{s}_{hyb})^2$. Here $\zeta$ depends on the geometry of the mode and is of the order of 1. 

\begin{figure}[H]
	\centering
	\includegraphics[scale=0.6]{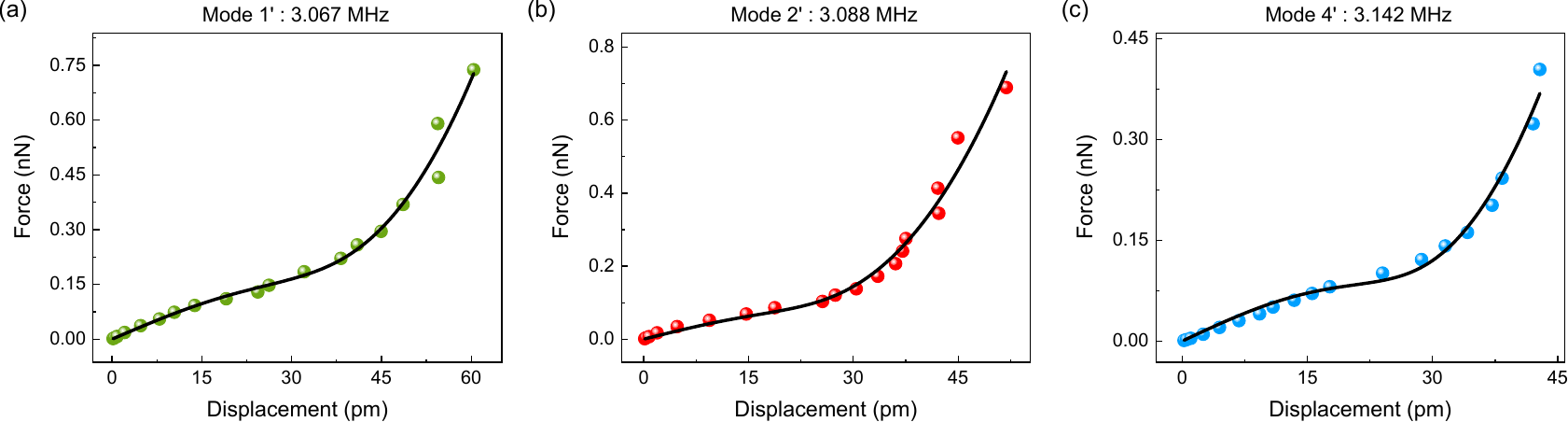}
	\caption{\textbf{Nonlinear response of hybrid SiNx modes:} \textbf{(a, b, c)} Force vs peak displacement plot of hybrid modes $1'$, $2'$ and $4'$ respectively, when coupled to graphene and probed on SiNx. SiNx modes experience a capacitive force due to ac gate voltage and a dominant backaction force from the graphene.} 
\end{figure}
The value of $\beta^{s}_{hyb1}$ for hybrid mode $1^\prime$ extracted from fitting result (Fig. S.9) is $5.6(\pm0.5)\times 10^{21}$ $\rm N/m^3$.
Similarly for hybrid mode $2^{\prime}$, $\beta^{s}_{hyb2}=8.0(\pm0.8)\times 10^{21}$ $\rm  N/m^3$ and for hybrid mode $4^\prime$, $\beta^{s}_{hyb4}= 7.6(\pm0.2)\times10^{21}$ $\rm  N/m^3$ respectively. One can also estimate the nonlinear coefficient of the hybrid SiNx modes using critical displacement of that mode~\cite{Zwickl-08}. It is given by:

\begin{equation}
\beta^{\rm s}_{hyb}=1.54\frac{m_s\omega_s^2}{Q_s {x^{s, cr}_{hyb}}^2}
\label{eqn:6}
\end{equation} 

For hybrid mode $1^{\prime}$, $x^{s,cr}_{hyb} = 38.2$ pm and $Q_s=1316.5$ result in $\beta^{\rm s}_{hyb1}= 7.1(\pm1.6)\times 10^{21}$ $\rm  N/m^3$. Similarly for hybrid mode $2^{\prime}$, $\beta^{\rm s}_{hyb2}=7.6(\pm1.5)\times10^{21}$ $ \rm  N/m^3$ and for hybrid mode $4^{\prime}$, $\beta^{\rm s}_{hyb4}=7.1(\pm1.2)\times10^{21}$ $ \rm  N/m^3$. It is remarkable to note that the hybrid modes of SiNx are well described by a Duffing-oscillator model and therefore, induced SiNx nonlinearities can be effectively described by Duffing constants for hybrid modes. 


\section{S.4. \textit{Giant}, induced nonlinearity measured on SiNx surface}

Here we provide few technical justifications for our usage of the term \textit{giant nonlinearity} for Duffing constant as measured on SiNx surface of graphene-SiNx hybrid modes. Our justification is based on two estimates, all of which show orders of magnitude changes: (i) a comparison of nonlinear threshold of our hybrid modes to that of bare SiNx resonators without hybridization, as measured by different groups~\cite{Zwickl-08}, (ii) a comparison  of average thermal displacement and threshold displacement for the onset of nonlinearity for bare graphene, SiNx and hybrid SiNx modes show four orders of magnitude reduction in ratio for hybrid modes. We further discuss the importance of having induced and tunable nonlinearity of SiNx resonator modes.

\subsection{A. Comparison of nonlinear threshold displacements for bare and hybrid SiNx resonators:}

The threshold displacement corresponding to the bare SiNx $ (x^{s, cr}_{bare} $) is extracted using following relation:

\begin{equation}
x_{bare}^{s,cr}=x^{s,cr}_{hyb}\sqrt{\frac{\beta^{s}_{hyb}}{\beta^{s}_{bare}}}
\label{eqn:7}
\end{equation} 
where $\beta^{s}_{hyb}$ and $\beta^{s}_{bare}(=5\times10^{12}$ $\rm N/m^3)$ denotes hybrid and bare (following ref.~\cite{Zwickl-08}) Duffing constant of SiNx. For mode 1, $\beta^{s}_{hyb1}=7.1\times10^{21}$ $\rm N/m^3$ and $x_{hyb1}^s=38$ pm results in $x_{bare}^{s,cr}=1.2$ $\rm \mu m$. Similarly for mode 2, $x_{bare}^{s,cr}=1.4$ $\rm \mu m$ and for mode 3, $x_{bare}^{s,cr}=1.2$ $\rm \mu m$. One can therefore note that such estimated displacement for onset of nonlinearity is 5 order of magnitude larger than that of hybrid SiNx.

\subsection{B. Comparison of thermal displacement and displacement corresponding to nonlinear threshold:}
The ratio of nonlinear threshold ($x^{g,cr}_{bare}$) and thermal displacement ($x^{g}_{th}$) for bare graphene is ($4.6\times10^{-9}/1.04\times10^{-12}$=) $4.4\times10^3$. Using our parameters and results of ~ref.~\cite{Zwickl-08}, in case of bare SiNx, the ratios are $(1.4\times10^{-6}/24.5\times 10^{-15}=)5.9\times10^7$, $(1.2\times10^{-6}/23.7\times 10^{-15}=)5.0\times10^7$ and $(1.2\times10^{-6}/25.4\times 10^{-15}=)4.7\times10^7$ for mode 1, 2 and 3 respectively. 

However, for the same hybrid mode measured on graphene, ratio is ($1.0\times10^{-9}/1.04\times10^{-12}$=) $1.0\times10^3$ and ($1.2\times10^{-9}/1.04\times10^{-12}$=)$1.1\times10^3$, same order as that of bare graphene.

In case of SiNx hybrid modes the ratio drops by four orders of magnitude to $(38.2\times10^{-12}/24.5\times 10^{-15}=)1.6\times10^3$, $(30.4\times10^{-12}/23.7\times 10^{-15}=)1.3\times10^3$ and $(31.5\times10^{-12}/25.4\times 10^{-15}=)1.2\times10^3$.

\subsection{C. Relevance of induced nonlinearity of SiNx modes:}

SiNx resonators have shown significant promise of observing quantum mechanical behavior for high-Q mechanical resonators at room-temperature. However, one needs to engineer nonlinearty in such a quantum device, to make it useful. After all, fluctuations of a classical resonator in thermal state is similar in shape in phase space to that of fluctuations of a linear harmonic oscillator deep in the quantum regime, dominated by zero point motion. For the resonator to be useful for precision measurement, one requires to squeeze the fluctuations in one quadrature: this require nonlinear interactions. Similarly, for gate operations in information devices, it is necessary to have conditional switching and phase shifts, both of which require nonlinear interactions between modes.

\section{S.5. Theoretical model}

In this section, we analyze the model and find that the nonlinear system can be described by hybrid modes to some extent, akin to that of normal modes for a corresponding linear system.

\subsection{A. Coupled linear SiNx and nonlinear graphene resonator}
\noindent
Our model is based on coupled modes of graphene and SiNx resonators, denoted by 1-dimensional amplitudes $x_g$ and $x_s$, respectively and is described by the set of equations:
\begin{subequations}
	\begin{equation} 
	\ddot{x}_g= -\gamma^g_{\rm bare} \dot{x}_g -\frac{\eta^g_{\rm bare}}{m_g} {x_g}^2 \dot{x}_g -\frac{\beta^g_{\rm bare}}{m_g} {x_g}^3 -[\omega_g^2 +\epsilon_p \cos(\omega_p t)] x_g +\frac{\alpha}{m_g} x_s
	\end{equation}
	\begin{equation}
	\ddot{x}_s= -\gamma^s_{\rm bare} \dot{x}_s -\omega_s^2 x_s +\frac{\alpha}{m_s} x_g
	\end{equation}
\end{subequations}
where $\gamma^k_{\rm bare}$ and $\omega_k $ $(k=g,s)$ represent linear damping and frequency of graphene and SiNx modes. Nonlinearity of graphene is quantified with two parameters: nonlinear damping $\eta^g_{\rm bare}$ and a cubic nonlinear response, characterized by its Duffing coefficient $\beta^g_{\rm bare}$. The graphene mode is bilinearly coupled to a SiNx mode which is modeled by an effective interaction Hamiltonian, $H_{\rm int} = \alpha x_g x_s$, where $\alpha$ is a coupling constant. SiNx is considered to be a linear oscillator in the range of forcing that we apply in our experiments.

\subsection{B. Normal modes at low-amplitudes: probe on graphene and on SiNx resonators}

At low external forcing, one can ignore nonlinear terms and thereby define two normal modes $x_1$ and $x_2$. These modes extend over the entire device. However, we detect either on graphene ($x_g$) or on SiNx ($x_s$), which can be expressed as:

\begin{subequations}
	\begin{equation} 
	x_g=\frac{1}{2 \sqrt{\alpha/m_s}} \left(x_2 + x_{1}\right) = x^g_2 + x^g_1,
	\end{equation}
	and
	\begin{equation} 
	x_s=\frac{1}{2 \sqrt{\alpha/m_g}} \left(x_2 - x_1\right) =  x^s_2 - x^s_1
	\end{equation}
\end{subequations}

\begin{figure}[H]
	\centering
	\includegraphics[scale=0.1]{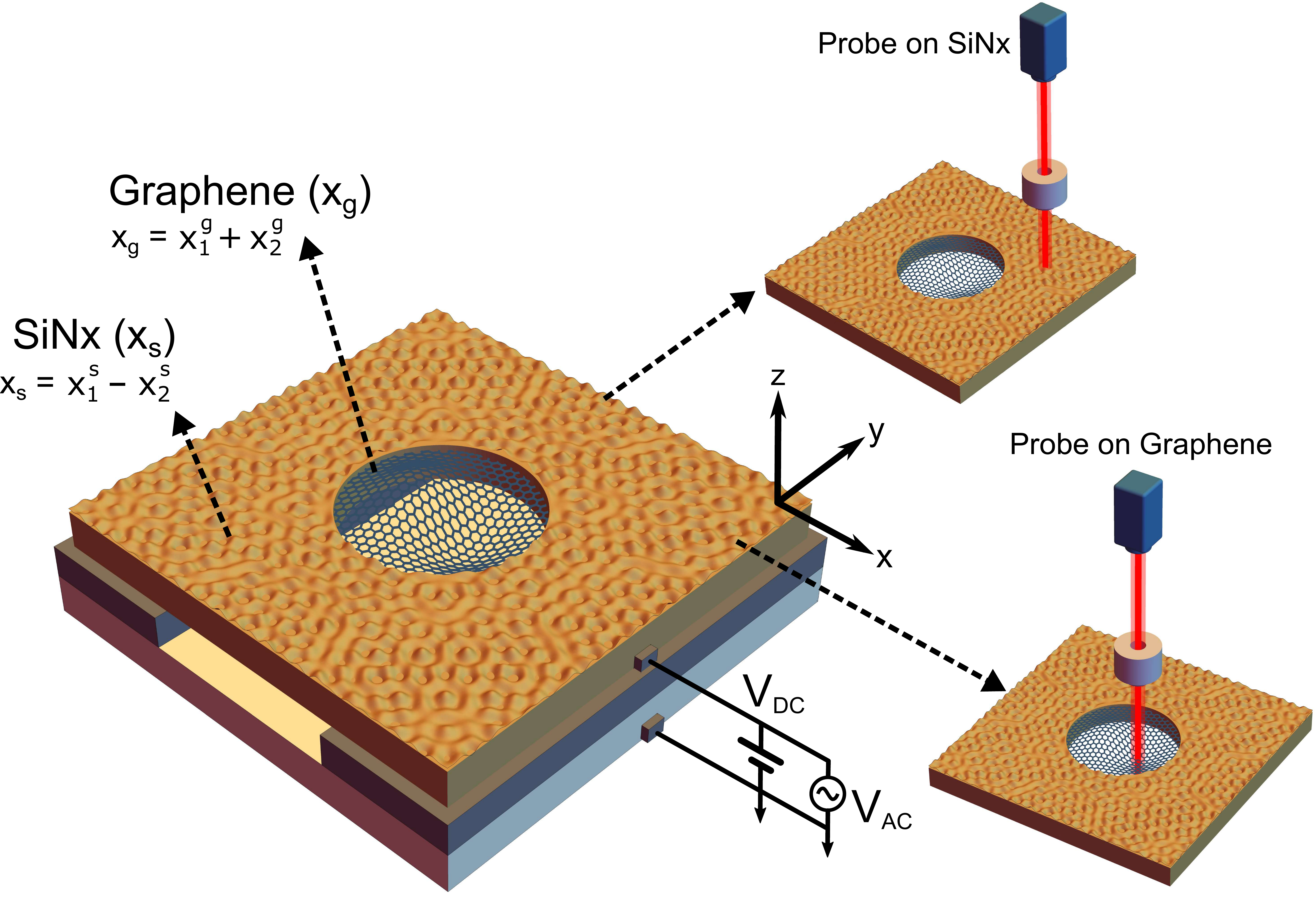}
	\caption{\textbf{Experimental schematics:} Probing the two resonators graphene and SiNx, which according to equations S.7a,b detects the motion of the coupled hybrid mode (i.e. mode 1 and mode 2).} 
\end{figure}

The detected amplitudes of normal mode $x_1$ (or $x_2$) on grpahene or SiNx are scaled by the ratio of square-root of respective masses ($m_s/m_g \sim 10^{4}$). As a result, amplitude of normal mode 1 (mode 2) on graphene i.e. $x^g_1$ ($x^g_2$) is two orders of magnitude larger than the amplitude of the same mode, $x^s_1$ ($x^s_2$), detected on SiNx surface. Accordingly, we have two Duffing constants for mode 1 (mode 2): $\beta^g_{\rm hyb 1}$ ($\beta^g_{\rm hyb 2}$) detected on graphene and $\beta^s_{\rm hyb 1}$ ($\beta^s_{\rm hyb 2}$) detected on SiNx.

\subsection{C. Perturbative estimation I: Difference in $\beta^{g/s}_{\rm hyb}$, measured on SiNx and graphene surfaces}

The difference in scales of Duffing constants measured on SiNx and on graphene surfaces, can be understood in the following way: it can be noted that the nonlinear forcing ($F_{NL}$) of a hybrid mode is uniform all along the spatial extent of the mode. However, since the hybrid mode for our device has physically two different kinds of oscillators with varying masses and surface areas, the force can be expressed as: $F_{NL} = \beta^g_{\rm hyb} x^3_g = \beta^s_{\rm hyb} x^3_s$, as measured on graphene ($x_g$) or on SiNx ($x_s $). For a forcing $F_0$ and assuming a steady state amplitude of $x_{g,s}\sim F_0Q_{g,s}/(m_{g,s}\omega^2_0)$ for graphene and SiNx, leads to an approximate ratio of the measured Duffing coefficients $\beta^s_{{\rm hyb}}/\beta^g_{{\rm hyb}}\sim(m_sQ_g/m_gQ_s)^3\sim 10^9$. This is in accordance with our measured values of $\beta^{g,s}_{{\rm hyb}}$ on graphene and on SiNx and provides a simple explanation of the giant nonlinearity measured on SiNx resonator surface. 

\subsection{D. Perturbative estimation II: effective nonlinearity}
To get an estimate of effective scaling of induced Duffing constant of SiNx hybrid mode, $\beta^s_{\rm hyb}$ to that of graphene's mass ($m_g$), bare Duffing constant, $\beta^g_{\rm bare}$ and coupling $\alpha$, perturbatively, let us consider the following simplified equations:
\begin{subequations}
	\begin{equation}
	\ddot{x}_g+\omega^2_0x_g+\frac{\beta^g_{\rm bare}}{m_g} x_g^3-\alpha_g x_s=0\,,
	\end{equation}
	
	\begin{equation}
	\ddot{x}_s+\omega_0^2x_s-\alpha_sx_g=0\,.
	\end{equation}
\end{subequations}
where  $\alpha_i = \alpha/m_i$, ($i=g,s$) and damping is ignored.

For uncoupled graphene mode ($\alpha_g = 0$) and assuming $\beta^g_{\rm bare}x_g^3\ll1$, standard perturbation methods yields a (zeroth-order) solution of the form:
\begin{equation}
x^{(0)}_g=A_g\cos\left[\left(\omega_0+\beta^g_{\rm bare}\frac{3A_g^2}{8\omega_0 m_g}\right)t\right]-\frac{\beta^g_{\rm bare} A_g^3}{32\omega_0^2 m_g}\left(\cos\omega_0t-\cos3\omega_0 t\right).
\label{eqn:14}
\end{equation}
where $A_g$ is a constant set by initial conditions.
Substituting this zeroth order expression of $x^{(0)}_g$ in equation S.9a, we arrive at
\begin{equation}
\ddot{x}_s+\omega_0^2x_s-\frac{\beta^s_{\rm hyb}}{m_s}x^3_s=B\cos\omega_0t,
\label{eqn:15}
\end{equation}
where, we have recognized $\cos\omega_0t$ as $x^{(0)}_s/A_s$ ($A_s$ being $x^{(0)}_s$ when $x_s$ and $x_g$ are uncoupled) and defined:
\begin{equation}
B\equiv\alpha_sA_g-\frac{\alpha_s\beta^g_{\rm bare} A_g^3}{8\omega_0^2m_g}\,,\\
\label{eqn:16}
\end{equation}
\begin{equation}
\beta^s_{\rm hyb}\equiv\frac{\alpha_s\beta^g_{\rm bare} A_g^3}{8\omega_0^2m_gA_s^3}\,.
\label{eqn:17}
\end{equation}

Here, in the definition of $B$, we have ignored frequency correction.

Substituting the values of $A_{g,s}=F_0 Q_{g,s}/m_{g,s}\omega_0^2$ in above expression cancels the common forcing ($F_0$) yielding an expression that depends only on the system parameters:

\begin{equation}
\beta^s_{\rm hyb}\equiv\frac{\alpha \beta^g_{\rm bare} m_s^3 Q_g^3}{8\omega_0^2 m_g^4 Q_s^3}
\end{equation}

where $\alpha$ is the coupling strength between two resonators, $\beta_{bare}^g$ is the Duffing constant of the bare grapene mode, $m_s (m_g)$ is the mass of  SiNx (graphene) resonator, $\omega_0$ is the on-resonant frequency of both the resonators, and $Q_s (Q_g)$ is the quantity factor of SiNx (graphene) modes. 

We further note that the sign of $\alpha$ determines whether the SiNx is effectively a soft or a hard nonlinear oscillator. From the expression of $\beta^s_{\rm hyb}$, one can then express an effective scaling as:
\begin{eqnarray}
\beta^s_{\rm hyb}\propto \frac{\alpha \beta^g_{\rm bare}m_s^3}{m_g^4}\,.
\label{eqn:18}
\end{eqnarray}

%
%
%

\newpage

\section{S.6. Frequency comb I: estimating nonlinear coefficients}

\noindent

We first develop a numerical model that reproduce the experimental observation of the frequency comb. Simulations results thereby give us estimate of $\beta^{g}_{\rm bare}$ and $\eta^{g}_{\rm bare}$. Next, we develop a general methodology to estimate nonlinear coefficients from measured experimental spectra on a general resonator surface. We finally apply the methodology to estimate $\beta^{k}_{\rm hyb }$ and $\eta^{k}_{\rm hyb}$ ($k= g,s$), as measured on graphene or SiNx surface. 

\subsection{A. Estimating nonlinear coefficient from simulated spectra} 

\noindent
The parameters used in numerical simulation of equations S.7a,b were extracted by fitting the Brownian spectrum (Fig. S.2) of graphene (equation S.2) and are listed in Table I. By varying the free parameters i.e. $\beta^{g}_{\rm bare}$ and $\eta^{g}_{\rm bare}$, we carefully calibrate and match the spectra in the instability region, where the comb is generated. The flow diagram in Fig. S.11 describes the methodology of nonlinearity estimation. 

\begin{figure}[H]
	\centering
	\includegraphics[scale=0.35]{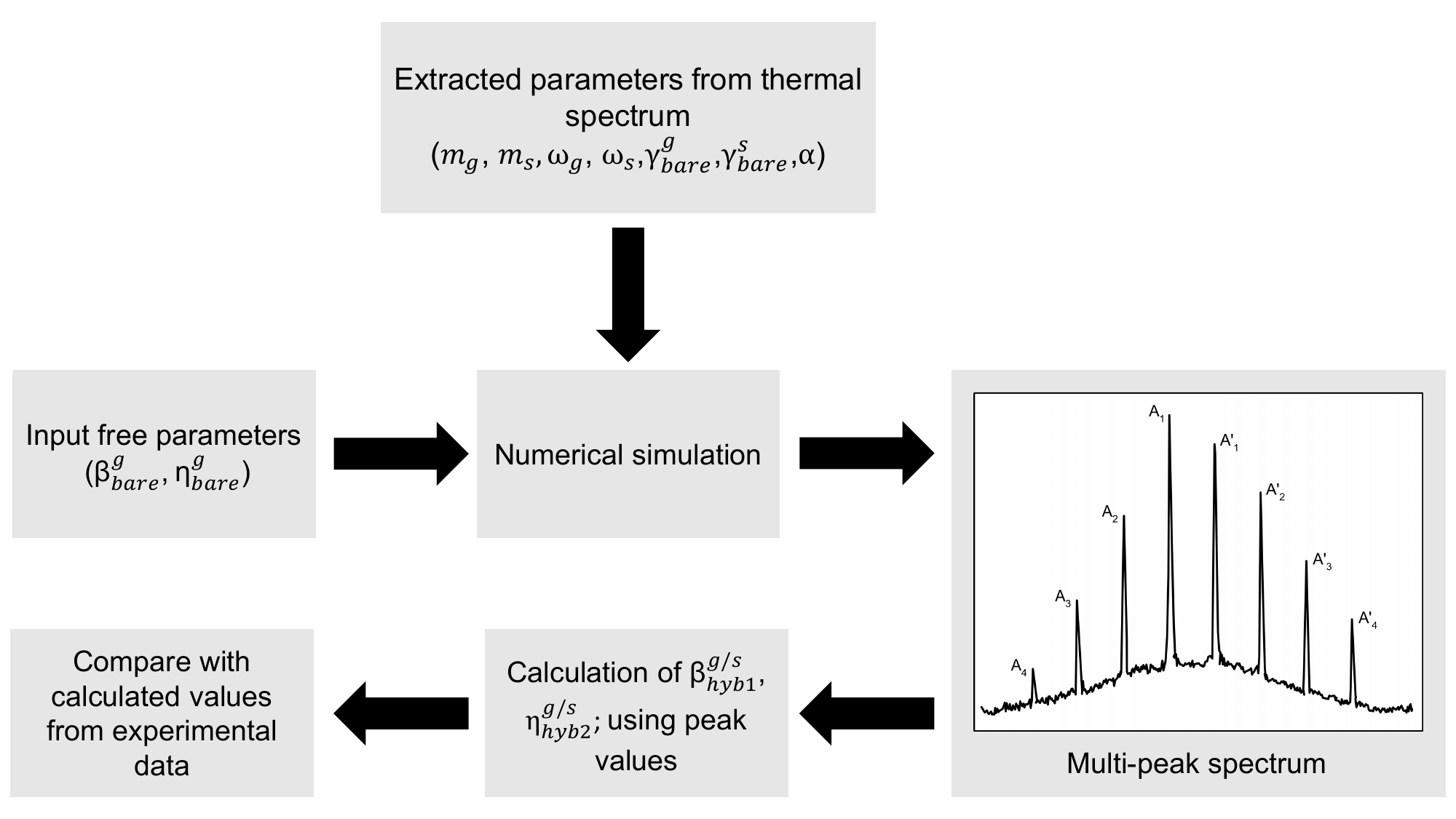}
	\caption{\textbf{Nonlinearity estimation from simulation:} The flow diagram represents our method to estimate the nonlinear coefficients $\beta^{g}_{\rm bare}$ and  $\eta^{g}_{\rm bare}$.} 
\end{figure}

\begin{table}[ht]
	\caption{Parameters for numerical simulation} 
	\centering 
	\begin{tabular}{c c c c} 
		\hline\hline 
		Parameter & unit & Fig. 2c, main text & Fig. 3c, main text \\ [0.5ex] 
		\hline 
		$m_g$ & kg &$10\times0.625\times10^{-16}$ & $10\times0.625\times10^{-16}$ \\ 
		$m_s$ & kg & $2.38\times10^{-11}$ & $2.38\times10^{-11}$ \\ 
		$\omega_g$& $s^{-1}$ & $2\pi\times2.864 \times 10^6$  & $2\pi\times3.005 \times 10^6$\\
		$\omega_s$& $s^{-1}$ & $2\pi \times 2.866\times10^{6}$ & $ 2\pi\times3.007\times 10^{6}$\\ 
		$\gamma^g_{bare}$ & $s^{-1}$ & $2\pi\times25050$ &$2\pi\times8045.8$\\
		$\gamma^s_{bare}$ & $s^{-1}$ & $2\pi\times744$ & $2\pi\times496.4$\\
		$\alpha$ &  $ \rm kg s^{-2}$ & $ 4\pi^2 \times 2.328\times10^{-3}$& $4\pi^2 \times 3.4\times 10^{-3}$ \\
		$\beta^{g}_{\rm bare}$ & $\rm N/m^3$  & $5.8\times 10^{13}$  & $1.07\times 10^{12}$\\
		$\eta^{g}_{\rm bare}$ & $\rm Ns/m^3$ &  $9.8\times 10^6$ & $7.5\times 10^{4}$\\
		\hline 
	\end{tabular}
	\label{table:nonlin} 
\end{table}

Numerically, we observe that the asymmetry in the envelop of the generated comb increases when nonlinear coefficient ($\beta^{g}_{\rm bare}$) is increased (Fig. S.12a) while the slope of the envelop changes with non-linear damping coefficient ($\eta^{g}_{\rm bare}$) (Fig. S.12b). The overall asymmetric fan-like shape of the generated comb is therefore a result of interplay between nonlinear damping and Duffing nonlinearity.

\begin{figure}[H]
	\centering
	\includegraphics[scale=0.13]{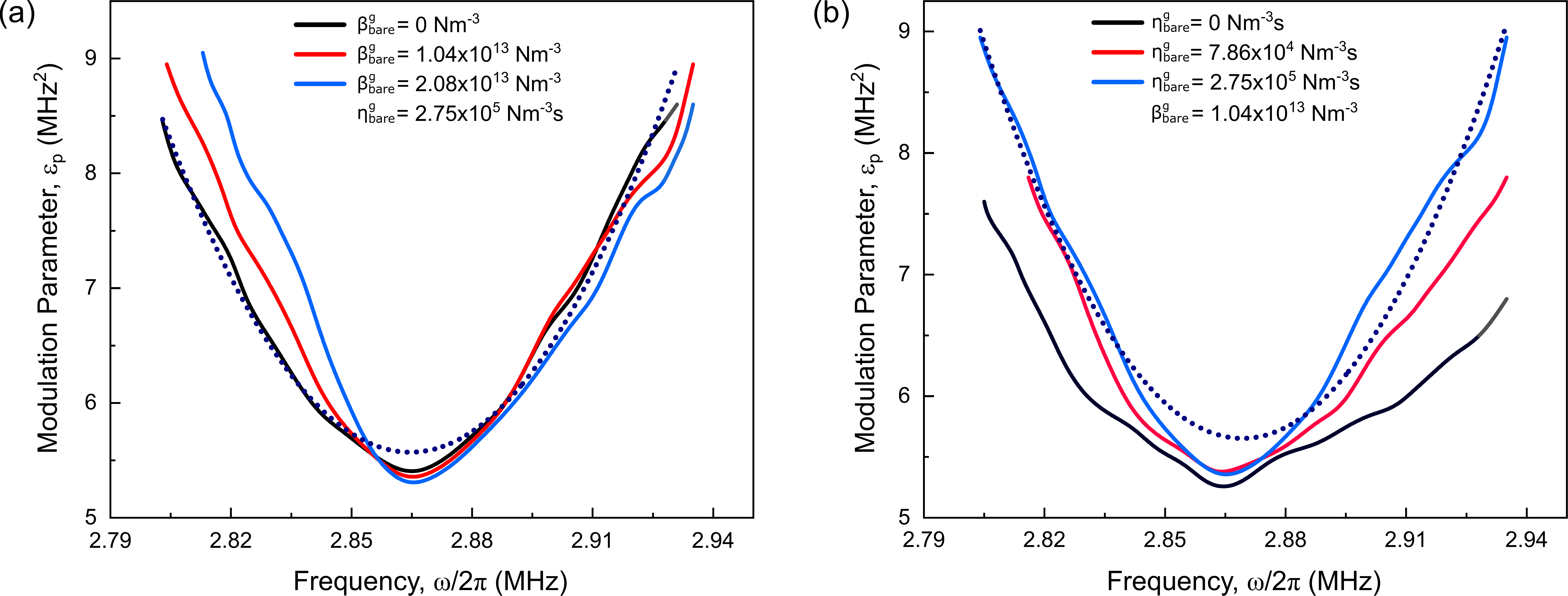}
	\caption{\textbf{Interplay of  Duffing constant and nonlinear damping:} The simulated profile of the instability region depends on the values of  $\beta^{g}_{\rm bare}$ and  $\eta^{g}_{\rm bare}$, the curves illustrate the envelop of the instability region in accordance with Fig. 3a, main text. \textbf{(a)} For a fixed  $\eta^{g}_{\rm bare}$ the asymmetry of the profile increases with increasing  $\beta^{g}_{\rm bare}$ values.  \textbf{(b)} For a fixed  $\beta^{g}_{\rm bare}$ value, the simulated profile becomes narrower with increasing  $\eta^{g}_{\rm bare}$ values. } 
\end{figure}

\begin{figure}[H]
	\centering
	\includegraphics[scale=0.4]{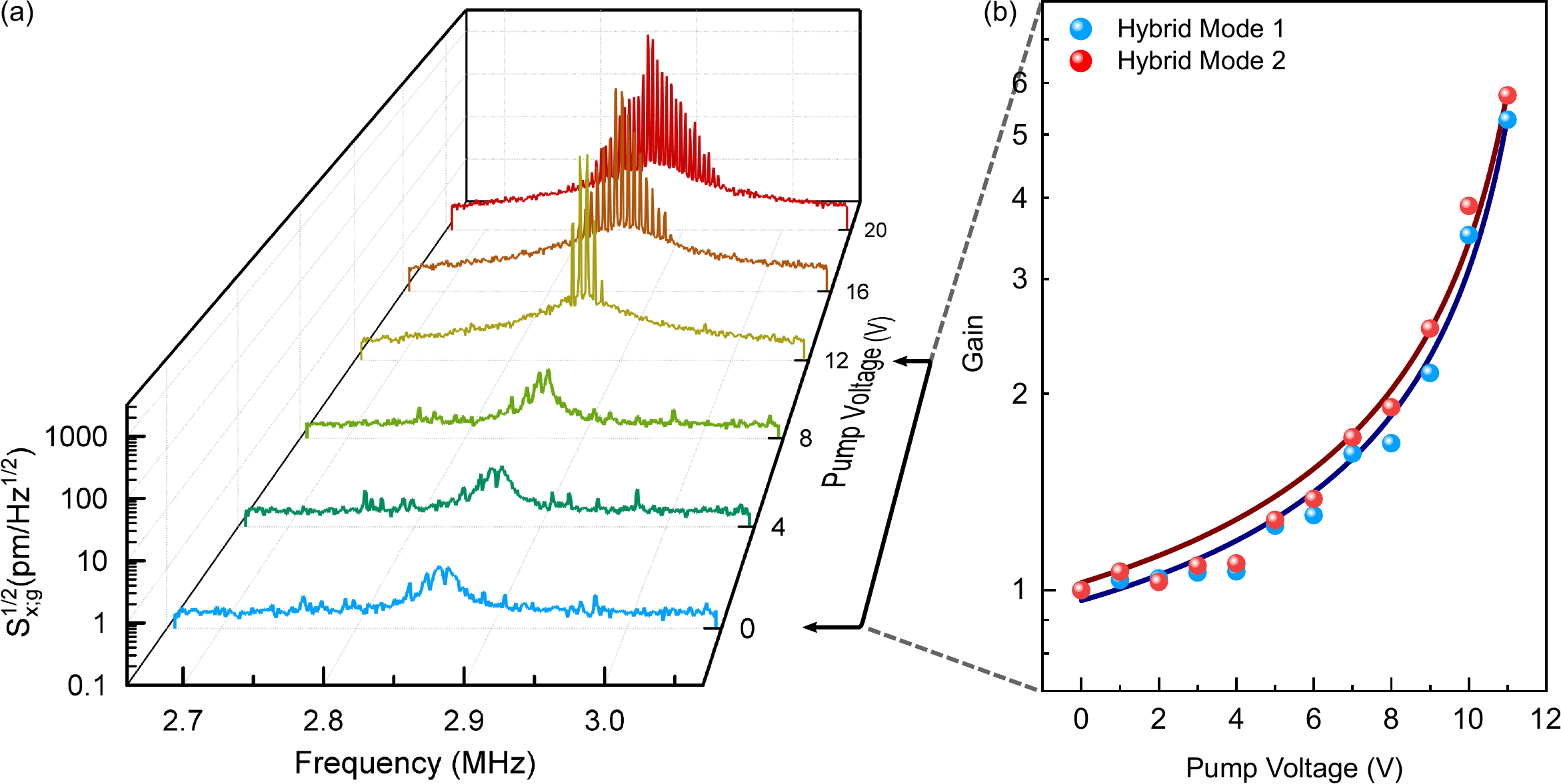}
	\caption{\textbf{Multimode spectrum on graphene:} \textbf{(a)} Selected cross-sectional plots of graphene-SiNx hybrid on graphene as a function of pump voltage from Fig. 2b, main text. \textbf{(b)} Plot of gain as a function of pump voltage for the hybrid modes up to the critical voltage, the fitting curves were referred from ref. ~\cite{Singh18}. 
	} 
\end{figure} 

\subsection{B. Methodology of nonlinearity estimation}

Here we describe the methodology we use to estimate nonlinear coefficients from observed frequency combs on graphene and SiNx surfaces.
When modes are driven at twice the resonance frequency, we observe parametric gain (Fig. S.13) in both the hybrid modes below a threshold pump voltage. Above threshold, in the self-oscillation regime, we observe mixing of modes. We attribute this mixing to nonlinearity in the system. Using amplitudes of newly generated modes, we estimate the corresponding nonlinear coefficients.

In particular, starting with amplitudes of four modes of frequency comb  to be $A_1$, $A'_1$, $A_2$ and $A'_2$, such that the corresponding displacement (measured on graphene or SiNx surface) can be expressed as:

\begin{equation}
x_k=(A_1 e^{-i\Delta t}+ A'_1 e^{i\Delta t}+A_2 e^{-i3\Delta t}+A'_2 e^{i3\Delta t})e^{i\omega_0 t}
\label{eqn:1}
\end{equation}

\noindent
where $k=g,s$ and $n\Delta$ $(n=1,3)$ is the separation of modes from the central frequency, $\omega_0$.
Combining equation S.16 with equation S.7, terms corresponding to Duffing nonlinearity and nonlinear damping can be expressed as:

\begin{multline}
\beta^k_{hyb} \left\langle x_k^2\right\rangle x_k+\eta^k_{hyb} \left\langle x^2\right\rangle \dot{x_k}=\\
\beta^k_{hyb} (A_1 e^{-i\Delta t}
+ A'_1 e^{i\Delta t}+A_2 e^{-i3\Delta t}+
A'_2 e^{i3\Delta t})^3e^{i\omega_0 t}\\
+i\eta^k_{hyb}(A_1 e^{-i\Delta t}+A'_1 e^{i\Delta t}+A_2 e^{-i3\Delta t}+
A'_2 e^{i3\Delta t})^2\\\{(\omega_0-\Delta) A_1 e^{-i\Delta t}+(\omega_0+\Delta) A'_1e^{i\Delta t}\}e^{i\omega_0 t}
\label{eqn:20} 
\end{multline}


\noindent
Using rotating wave approximation and collecting the terms corresponding to modes at $\mp3\Delta$ from central frequency, one gets:

%
\begin{subequations}
	\begin{equation}
	(\beta^k_{hyb})^2+(\omega_0-\Delta)^2(\eta^k_{hyb})^2=36\bigg(\frac{A_2}{A_1^3}\bigg)^2 \Delta^2\omega_0^2 m_g^2
	\label{eqn:23}
	\end{equation}
	
	\begin{equation}
	(\beta^k_{hyb})^2+(\omega_0+\Delta)^2(\eta^k_{hyb})^2=36\bigg(\frac{A'_2}{{A'}_1^3}\bigg)^2 \Delta^2\omega_0^2 m_g^2
	\label{eqn:24}
	\end{equation}
\end{subequations}

\noindent
Similarly, collecting terms corresponding to $\mp5\Delta$ from equation S.7 and squaring, yields

%

\begin{subequations}
	\begin{equation}
	9(\beta^k_{hyb})^2+(3\omega_0-5\Delta)^2(\eta^k_{hyb})^2=100\bigg\{\frac{A_3^2}{(A_1^2A_2+A_1^{'}A_2^2)^2}\bigg\}\Delta^2\omega_0^2 m_g^2
	\label{eqn:25}
	\end{equation}
	\begin{equation}
	9(\beta^k_{hyb})^2+(3\omega_0+5\Delta)^2(\eta^k_{hyb})^2=100\bigg\{\frac{{A'}_3^2}{({A'}_1^2{A'}_2+A_1{A'}_2^2)^2}\bigg\}\Delta^2\omega_0^2 m_g^2
	\label{eqn:26}
	\end{equation}
\end{subequations}
where $A_3$ and  $A'_3$ depicts amplitude of newly generated modes, emerging at $\mp5\Delta$ from central frequency.
\noindent
Solving equation S.18a and S.19a for the nonlinear coefficients, we finally get:
\begin{equation}
\beta^k_{hyb2}=\frac{\omega_0\Delta m_g}{\sqrt{12 \omega_0\Delta}}\bigg[100(\omega_0^2-2\omega_0\Delta)\frac{A_3^2}{(A_1^2A_2+{A'}_1A_2^2)^2}\\-
36(9\omega_0^2-30\omega_0\Delta)\bigg(\frac{A_2}{A_1^3}\bigg)^2\bigg]^{1/2}
\label{eqn:27}
\end{equation}
and,
\begin{equation}
\eta^k_{hyb2}=\frac{\omega_0\Delta m_g}{\sqrt{12 \omega_0\Delta}} \bigg[-100\frac{A_3^2}{(A_1^2A_2+{A'}_1A_2^2)^2}\\+324\bigg(\frac{A_2}{A_1^3}\bigg)^2\bigg]^{1/2}
\label{eqn:28}
\end{equation}
\noindent
where $\beta^k_{hyb2}$ and $\eta^k_{hyb2}$ are nonlinear damping and Duffing nonlinear coefficient of the left ($\omega_1$) hybrid mode. Similarly solving equation S.18b and S.18b for the right ($\omega_2$) hybrid mode we get:
\begin{equation}
\beta^k_{hyb1}=\frac{\omega_0\Delta m_g}{\sqrt{12 \omega_0\Delta}}\bigg[-100(\omega_0^2+2\omega_0\Delta)\frac{A_3^2}{(A_1^2A_2+{A'}_1A_2^2)^2}\\+36 (9\omega_0^2+30\omega_0\Delta)\bigg(\frac{A_2}{A_1^3}\bigg)^2\bigg]^{1/2}
\end{equation}
\begin{equation}
\eta^k_{hyb1}=\frac{\omega_0\Delta m_g}{\sqrt{12 \omega_0\Delta}}\bigg[100\frac{A_3^2}{(A_1^2A_2+{A'}_1A_2^2)^2}\\-324\bigg(\frac{A_2}{A_1^3}\bigg)^2\bigg]^{1/2}
\end{equation}
This method gives an estimate of the nonlinear coefficients of coupled hybrid graphene-SiNx mode  from spectral measurements.

\subsection{C. Application I: Estimating nonlinear coefficients on graphene resonator surface}

Based on the methodology discussed in appendix, we estimate values of $\beta^{g}_{hyb(2,1)}$ and $\eta^{g}_{hyb(2,1)}$ (corresponding to experimental data of Fig. 2b, main text) for every pump voltage above threshold (Fig. S.14). 

\begin{figure}[H]
	\centering
	\includegraphics[scale=1.0]{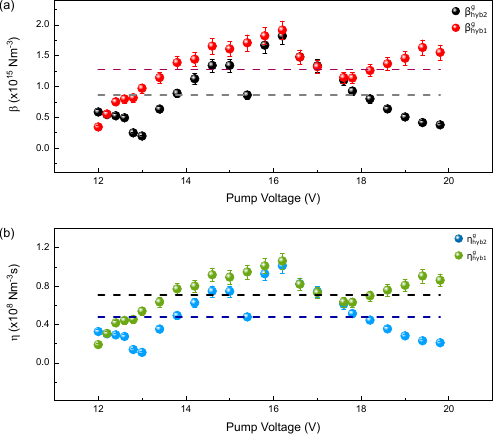}
	\caption{\textbf{Estimated nonlinear parameters:} Following our nonlinear coefficient estimation scheme, we determine \textbf{(a)} nonlinear coefficient, $\beta^g_{hyb(2,1)}$ and \textbf{(b)} nonlinear damping, $\eta^g_{hyb(2,1)}$ values for Fig. 2b, main text. The dashed lines in the plots indicate the average value of the extracted parameters.} 
\end{figure}

\noindent
Corresponding values of $\beta^{g}_{hyb(2,1)}$ and $\eta^{g}_{hyb(2,1)}$ as obtained from numerical simulation (corresponding to Fig. 2c, main text) are also plotted with pump voltage (Fig. S.15). There is an overall agreement between the experimental observations and numerical simulations.  
\begin{figure}[H]
	\centering
	\includegraphics[scale=1.0]{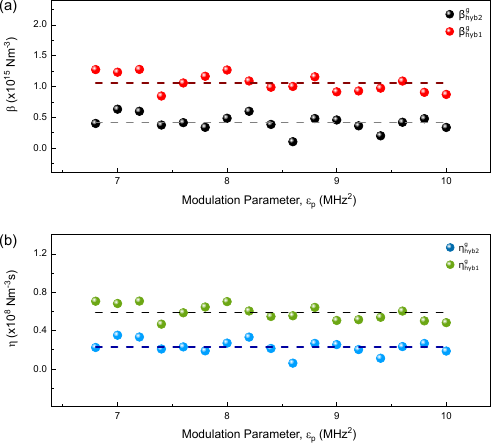}
	\caption{\textbf{Estimated nonlinear parameters from simulation:} We determine \textbf{(a)} nonlinear coefficient, $\beta^g_{hyb(2,1)}$ and \textbf{(b)} nonlinear damping, $\eta^g_{hyb(2,1)}$  from Fig. 2c, main text. The values of nonlinear coefficients remain fairly constant with the pump voltage. The dashed lines indicate the average value of calculated parameters.} 
\end{figure} 

This method gives an estimate of the nonlinear coefficients of coupled hybrid graphene-SiNx mode from spectral measurements.


\



\subsection{D. Application II: Estimating Duffing constant on SiNx resonator}

\noindent
We have already established graphene to be nonlinear resonator with many intriguing properties in parametric regime, which we expect to observe in SiNx at resonance with graphene. However due to huge mass of SiNx, the nonlinear signature is not easily detectable when probed on SiNx. The signature of parametrically driven graphene-Silicon Nitride hybrid mode is observed with small number of new generated modes (Fig. S.16). Looking at the asymmetry we can conclude about large nonlinearity.
\begin{figure}[H]
	\centering
	\includegraphics[scale=0.5]{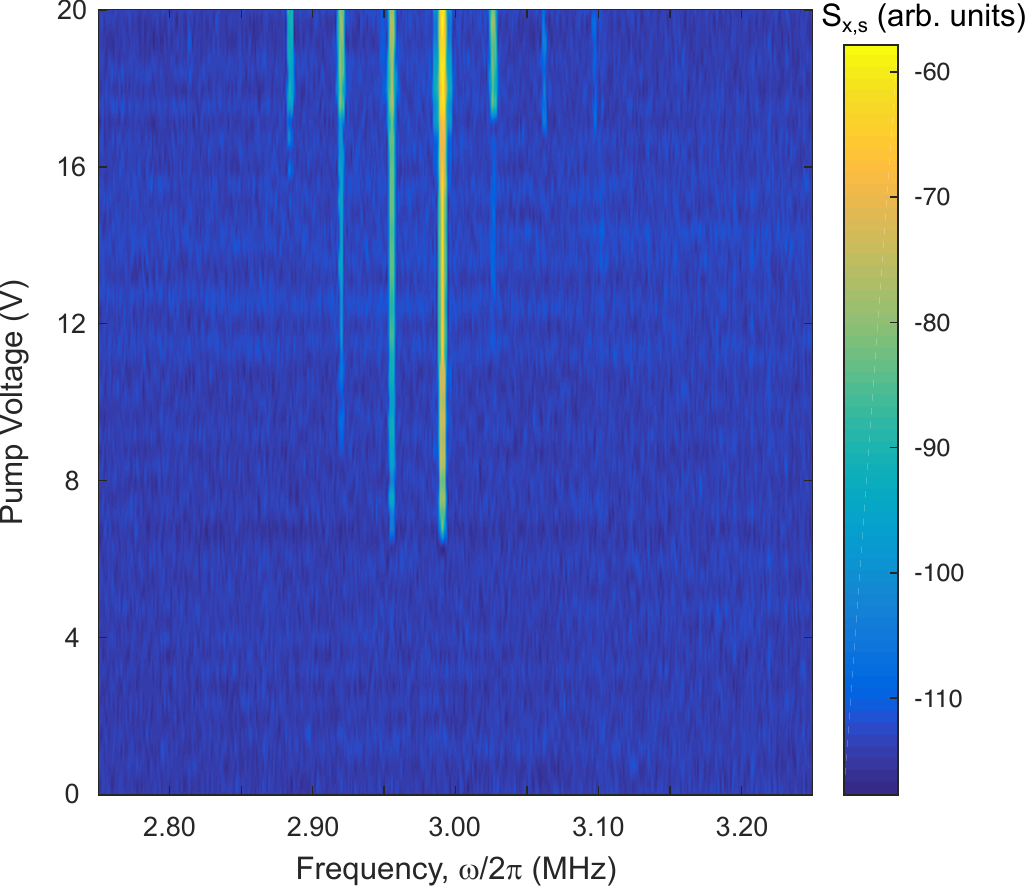}
	\caption{\textbf{Frequency comb on SiNx surface:} Induced multi-mode spectrum on SiNx as a function of pump voltage.} 
\end{figure}

\begin{figure}[H]
	\centering
	\includegraphics[scale=0.86]{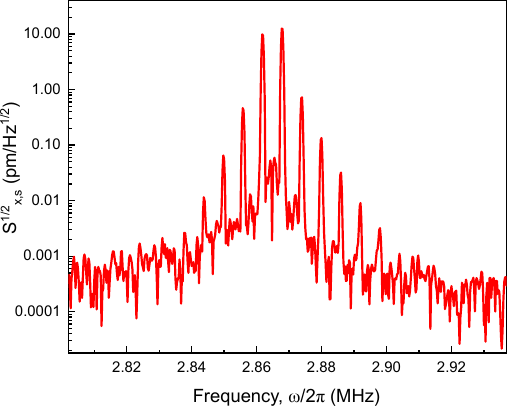}
	\caption{\textbf{Simulated multimode spectrum on SiNx}: Simulation of induced multi-mode spectrum in SiNx.} 
\end{figure}

\noindent

We simulate using equations S.7a,b, where SiNx is treated as a linear resonator, i.e., $\beta_s,\eta_s=0$ and observe multi-mode generation in SiNx spectrum (Fig. S.17), further validating our observation. The nonlinear coefficients estimated from simulation plots (Fig. S.17) using equations S.20-22 and S.23 turns out to be, $\beta^s_{hyb2}(\beta^s_{hyb1})$ = $ 4.9\times 10^{23}(4.7\times10^{23})$ $\rm N/m^3$ and $\eta^s_{hyb2}(\eta^s_{hyb1})$ = $ 2.7\times 10^{16}(2.6\times10^{16})$ $\rm  Ns/m^3 $, in harmony with the experimentally measured values.

\def\bibsection{\section*{\refname}}

\end{widetext}

\end{document}